\pdfoutput=1

\documentclass[11pt]{article}

\usepackage[final]{acl}

\usepackage{times}
\usepackage{latexsym}

\usepackage[T1]{fontenc}

\usepackage[utf8]{inputenc}

\usepackage{microtype}

\usepackage{inconsolata}

\usepackage{graphicx}

\usepackage{subcaption}
\usepackage{marvosym}
\usepackage{tikz}
\newcommand*\circled[1]{\tikz[baseline=(char.base)]{
            \node[shape=circle,draw,inner sep=0.7pt] (char) {#1};}}

\usepackage{algorithm}
\usepackage{algorithmic}
\usepackage{comment}
\usepackage{natbib}
\usepackage{wrapfig}
\usepackage{circledsteps}
\usepackage{multirow}
\usepackage{subcaption}
\usepackage{amsmath,amssymb,amsfonts}
\usepackage{algorithmic}
\usepackage{graphicx}
\usepackage{textcomp}
\usepackage{xcolor}

\newcommand{\tooluns}{\textsc{UnsuperBinTrans}}
\newcommand{\toolmal}{\textsc{FlowMalTrans}}
\newcommand{\tooluni}{\textsc{UniMap}}
\newcommand{\toolcross}{\textsc{CrossIns2Vec}}

\newcommand{\aaf}{\vspace*{-6pt}}
\newcommand{\af}{\vspace*{-3pt}}

\newcommand{\todo}[1]{\textcolor{black}{#1}}

\title{FlowMalTrans: Unsupervised Binary Code Translation for Malware Detection Using Flow-Adapter Architecture}

\author{Minghao Hu \\
  George Mason University\\
  \texttt{mhu20@gmu.edu} \\\And
  Junzhe Wang \\
  George Mason University \\
  \texttt{jwang69@gmu.edu} \\ \And
  Weisen Zhao \\
  George Mason University \\ 
  \texttt{wzhao9@gmu.edu} \\ 
  \AND
  Qiang Zeng \\
  George Mason University \\
  \texttt{zeng@gmu.edu} \\ \And
  Lannan Luo \textsuperscript{\Letter} \\
  George Mason University \\
  \texttt{lluo4@gmu.edu} \\}

\begin{document}
\maketitle
\begin{abstract}
Applying deep learning to malware detection has drawn great attention due to its notable performance. 
As malware spreads across a wide range of Instruction Set Architectures (ISAs), it is important to extend malware detection capacity to multiple ISAs.
However, training a deep learning-based malware detection model usually requires a large number of labeled malware samples.
The process of collecting and labeling sufficient malware samples to build datasets for each ISA is labor-intensive and time-consuming.
To reduce the burden of data collection, we propose to leverage the ideas of Neural Machine Translation (NMT) and Normalizing Flows (NFs) for malware detection. Specifically, when dealing with malware in a certain ISA, we translate it to an ISA with sufficient malware samples (like X86-64). This allows us to apply a model trained on one ISA to analyze malware from another ISA. 
We have implemented and evaluated the model on \emph{seven} ISAs.
The results demonstrate the high translation capability of our model, enabling superior malware detection across ISAs. \todo{We have made our code and dataset open-sourced available\footnote{\url{https://github.com/mhu16419/FlowMalTrans}}.}

\end{abstract}
\section{Introduction}
Malware is defined as software intended to damage computers or associated systems~\cite{preda2008semantics}. In recent years, many malware detection tools have been developed~\cite{xie2024mmpalm}, and their success largely depends on the underlying techniques. Signature-based detection identifies malware by matching patterns from known malware families~\cite{sathyanarayan2008signature}, but it often fails to detect altered or novel malware. In contrast, behavior-based detection analyzes program execution to identify suspicious behavior~\cite{liu2011behavior}, yet this method lacks scalability.

The use of deep learning methods like LSTM and CNNs in malware detection has garnered significant interest because of their strong performance~\cite{sewak2018investigation,gopinath2023comprehensive}.
However, there are two major challenges that hinder the broader adoption of deep learning in this domain.

\noindent \textbf{Challenges.} Malware infects a broad range of Instruction Set Architectures (ISAs). With the rise in cyberattacks on IoT devices~\cite{chen2025tracking,ma2023no,tsai2024anomaly}, malware has begun targeting an increasing number of ISAs~\cite{davanian2022malnet,caviglione2020tight}. Hence, it is essential to enable malware detection across ISAs. Currently, there are numerous ISAs available~\cite{wang-security23}. Due to this wide variety, training one model for each ISA demands substantial time and effort. In addition, deep learning models rely heavily on having enough malware samples during training, which is problematic for \emph{low-resource} ISAs where such data is limited.

\noindent \textbf{Our Approach.} Malware is generally closed-source whose original source code is typically inaccessible. What is available is the binary form of the malware. Once disassembled, this binary can be represented in assembly language. Based on this observation, we propose leveraging Neural Machine Translation (NMT) techniques~\cite{artetxe2017unsupervised} to facilitate malware analysis.

When handling a binary in a given ISA (referred to as the \emph{source} ISA), we translate it to an ISA with rich malware samples, such as X86-64, which we refer to as the \emph{target} ISA. Once translated, we use a model trained on the \emph{target} ISA to test the translated code. This approach facilitates malware detection across multiple ISAs using a model trained exclusively on the target ISA, eliminating the need for extensive malware samples in other ISAs.

In order to capture ISA-specific syntax and semantic information, we leverage normalizing flows (NFs), a class of invertible transformations that  models complex distributions while maintaining tractability. Traditional translation models often struggle with capturing complex dependencies and handling the variation in data distributions. NFs address this challenge by modeling complex, high-dimensional distributions through invertible transformations, enabling more efficient learning of intricate relationships between source and target data. By doing so, we ensure a smooth and structured mapping between ISAs, leading to more precise and reliable translations. Furthermore, normalizing flows enable efficient sampling and density estimation, which improves the adaptability of our model to unseen binaries and enhances generalization across different architectures.



We design~\toolmal, which is a flow-adapter-based unsupervised binary code translation architecture. 
This design allows for robust feature alignment between different ISAs, improving translation fidelity and ultimately enhancing malware detection across ISAs.
To bridge the gap between the latent space of the source and target ISA, \toolmal\ learns an invertible transformation of binary representations from one ISA to another, facilitating robust feature alignment across different ISAs and enhancing translation fidelity.
\toolmal\ operates in a completely unsupervised manner, eliminating the need for parallel datasets. Notably, the training of \toolmal\ does not require any malware samples and relies only on binaries compiled from the abundance of opensource programs. \emph{Despite never encountering any malware samples during training, \toolmal\ is still capable of translating malware across ISAs with high translation quality}.



\noindent \textbf{Results.} We have implemented \toolmal, and evaluated its performance on \emph{seven} ISAs, including X86-64, i386, ARM32, ARM64, MIPS32, PPC32 and M68K. \toolmal\ achieves higher BLEU score than
the baseline method \tooluns~\cite{ahmad-luo-2023-unsupervised}.
Furthermore, we apply \toolmal\ to the malware detection task 
and achieve exceptional results, highlighting its superior translation quality, which enables highly effective malware detection across ISAs.
Below we highlight our contributions:
\begin{itemize}
\aaf\af
\item We propose \toolmal, a novel unsupervised approach to translate binaries across different ISAs. \toolmal\ addresses the data scarcity issue by enabling the detection of malware in low-resource ISAs using a model trained on the high-resource ISA. 
Importantly, the training of \toolmal\ does not require any malware samples, yet it is still capable of translating malware across ISAs and achieves high translation quality.
\aaf\af
\item We leverage normalizing flows (NFs)  to capture ISA-specific features. With NFs, \toolmal\ captures code semantics and and properly decodes  instruction sequences. 



\aaf\af
 \item The training of~\toolmal~does not require any malware samples, yet the model is still capable of translating malware across ISAs and achieves high translation quality.
 \aaf\af
\item We expand the evaluation to cover a \emph{broader range of ISAs} compared to prior work, showcasing \toolmal's wider applicability and improved performance. 

\aaf\af
\end{itemize}
\section{Related Work}

\subsection{Malware Detection}

\noindent \textbf{Signature-Based Detection.} Conventional malware detection methods primarily use signature-based approaches, working by recognizing malicious code patterns~\cite{sebastio2020optimizing,rohith2021comprehensive,sathyanarayan2008signature,behal2010signature}.
However, these methods are ineffective against newly developed or altered malware that modifies its code to avoid being identified.

\noindent \textbf{Behavior-Based Detection.}
This approach detects malware by examining program behavior during execution~\cite{liu2011behavior,burguera2011crowdroid,aslan2021intelligent,saracino2016madam}.
However, it may produce false negatives when malicious actions are not activated during observation.

\noindent \textbf{Deep Learning-Based Detection.}
Deep learning has been widely used for detecting malware. A range of models, including CNNs and LSTMs, have been utilized~\cite{sewak2018investigation,gopinath2023comprehensive,wang2024learning,he2023clur,lei2022uncertainty,zhang2024ai,zhao2025efficient}.
Nonetheless, these models generally demand substantial datasets of malware samples for training. Consequently, most existing methods concentrate on widely used ISAs like X86 and ARM, leaving \emph{low-resource} ISAs relatively unaddressed.

\subsection{{Normalizing Flows}}

Normalizing flows (NFs) are a family of generative models that transform a simple base distribution into a complex target distribution via a series of invertible transformations~\cite{ho2019flow++}. 
The key idea is to apply a sequence of differentiable mappings to a latent variable. 
Different types of NFs include coupling layers~\cite{dinh2014nice}, autoregressive flows~\cite{papamakarios2017masked}, and continuous normalizing flows~\cite{chen2018neural}. 

The flow adapter architecture has been explored in neural machine translation (NMT), such as variational NMT~\cite{calixto2019latent,ma2019flowseq,eikema2019auto}. Flow adapters leverage NFs to model complex posterior distributions in sequence-to-sequence models, allowing for more flexible latent representations and enhancing the robustness of translation~\cite{shu2020latent}. 

\subsection{Source Code Translation}

Source code translation converts source code in one language to another. Early work uses \emph{transpilers} or \emph{transcompilers}~\cite{andres2017transpiler,tripuramallu2024towards}, which rely on handcrafted rules. However, they produce unidiomatic translations that prove hard for human programmers to read. Another issue is incomplete feature support, resulting in improper translations.

Deep learning have introduced new approaches to source code translation~\cite{roziere2020unsupervised,weisz2021perfection,lachaux2020unsupervised}. However, since malware is closed-source,
\emph{source code translation is \textbf{inapplicable} for malware}.



\subsection{{Program Analysis-Based Binary Code Translation}}

Several approaches leverage program analysis techniques to convert binary code from one ISA to another. They can be broadly categorized into \emph{static analysis-based} and \emph{dynamic analysis-based}. Static analysis-based translation analyzes and translates the binary code before execution~\cite{shen2012llbt,cifuentes2000uqbt}. But ensuring all paths are accurately translated is challenging.

Dynamic analysis-based translation performs translation during execution~\cite{chernoff1998fx,ebcioglu2001dynamic}, but it requires sophisticated runtime environments.
Plus, both approaches face challenges related to \emph{architecture-specific features} and encounter difficulties in achieving accuracy, performance, and compatibility.

\noindent \textbf{Summary.} We propose an unsupervised binary code translation approach leveraging \textbf{\em deep learning techniques}. The state of the art, \tooluns~\cite{ahmad-luo-2023-unsupervised}, is the first and only existing work \emph{applying deep learning to binary code translation}. However, it has several limitations, which our model \toolmal\ aims to overcome (Section~\ref{subsec:limit-prior}). Our evaluation demonstrates that \toolmal\ outperforms \tooluns\ in binary code translation and achieves superior malware detection performance.

\section{Overview} \label{sec:overview}

\begin{figure*}[t]
    \centering
    \includegraphics[width=0.98\linewidth]{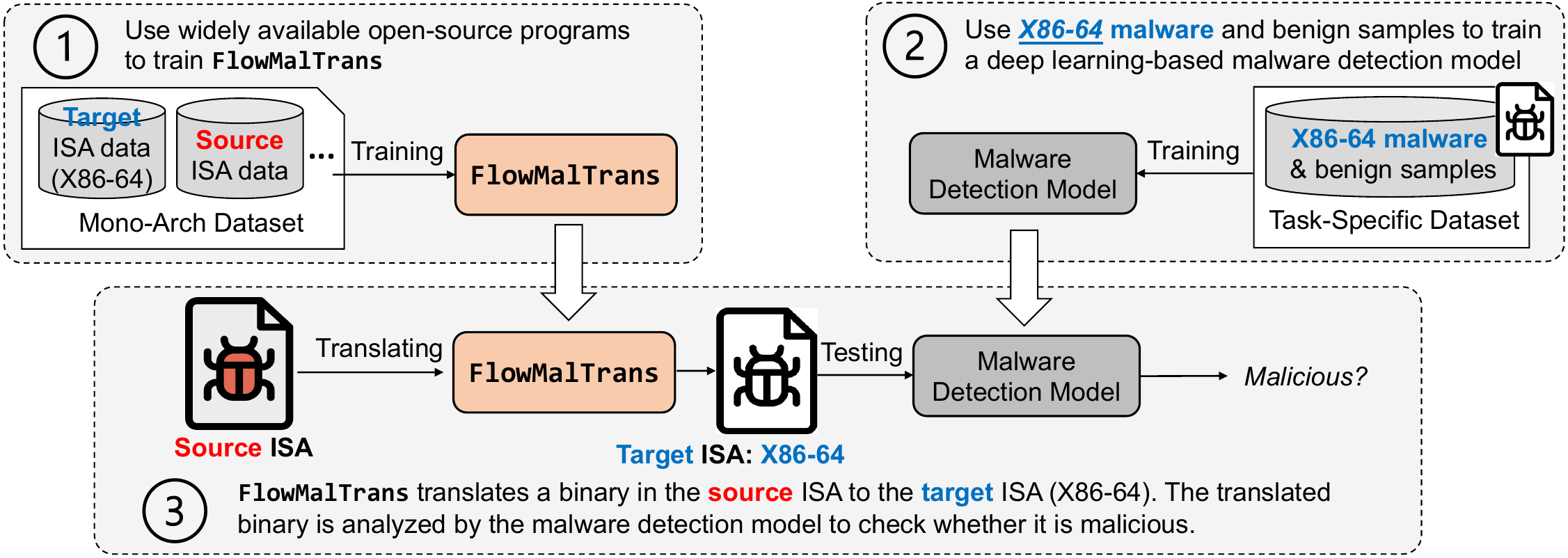}
    \caption{Applying \toolmal\ to detect malware in a source ISA using a malware detection model trained on the target ISA (e.g., X86-64).}
    \label{fig:motivation}
    \af
\end{figure*}

\subsection{Motivation}

Malware propagates over various instruction set architectures (ISAs), highlighting the importance of enabling malware detection across multiple ISAs. Nonetheless, training deep learning models generally depends on extensive datasets of malware samples. This requirement becomes more difficult to fulfill with the diversity of ISAs, as \emph{low-resource} ISAs frequently lack adequate malware samples. 

To alleviate the challenges of data collection and tackle the problem of limited data, we propose translating binaries from a \emph{low-resource} ISA to a \emph{high-resource} ISA.
Thus malware detection model trained on the \emph{high-resource} ISA can be applied to the translated binaries, helping to overcome the data scarcity faced by \emph{low-resource} ISAs.


\subsection{Limitations of the State of the Art} \label{subsec:limit-prior} 

\tooluns~\cite{ahmad-luo-2023-unsupervised} is the first and only existing work that applies deep learning techniques to binary code translation.
While InnerEye~\cite{zuo2018neural} applies NMT techniques for binary code similarity comparison, it  \emph{does not perform binary code translation across ISAs}. Instead, it employs two encoders from NMT models to generate embeddings for binary pairs.
\tooluns, however, has several limitations, which our model aims to overcome.


\noindent \textbf{Shared Encoder \emph{vs.} Separated Encoders.} \tooluns\ employs a shared encoder for multiple ISAs, which
limits the model's ability to capture ISA-specific syntax and semantics, leading to suboptimal translation. In contrast, \toolmal\ employs a separate encoder for each ISA, allowing each to learn specialized representations. 


\noindent \textbf{RNN \emph{vs.} Transformer.} \tooluns\ relies on RNNs, which suffer from vanishing gradients and difficulties in capturing long-range dependencies. In contrast, \toolmal\ leverages Transformer, which employ self-attention mechanisms to model long-range dependencies. 

\noindent \textbf{ISA-Agnostic \emph{vs.} ISA-Specific Latent Representations.} \tooluns\ learns  latent representations for binaries that are ISA-agnostic by abstracting away ISA-specific details. 
In contrast, \toolmal\ learns ISA-specific latent representations that preserve crucial architectural distinctions while maintaining alignment with a shared cross-ISA space. This is achieved by employing NFs. By transforming the source-ISA representations into the target-ISA representations via NFs, the decoder can leverage the representations to generate better aligned target-ISA binary code.

\noindent \textbf{Two ISAs \emph{vs.} Broader ISAs.} \tooluns\ is limited to only two ISAs (X86-64 and ARM32), whereas \toolmal\ extends its coverage, supporting seven ISAs.

\subsection{Model Applications} 

Figure~\ref{fig:motivation} shows the workflow of applying \toolmal\ to detect malware across ISAs.
In step \circled{1}, we train \toolmal\ to translate binaries from the source ISA to the target ISA (such as X86-64). The training dataset can be constructed using various available \emph{opensource} programs. 
\emph{The training of \toolmal\ does not require any malware samples}. It should be noted that malware is typically a closed-source program; thus \emph{cross-compilation that generates binaries across ISAs from source code does not apply to malware}.

In step \circled{2}, we train a deep learning model using a \emph{task-specific dataset} containing malware and benign samples in the target ISA. 

Finally, in step \circled{3}, when dealing with a binary in the source ISA, we use \toolmal\ to translate the binary to the target ISA and reuse the malware detection model trained on the target ISA to test the translated code for detecting malware. 

Note that due to the scarcity of malware samples in a low-resource ISA, directly training a robust malware detection model on such an ISA is challenging. Our approach overcomes this limitation through code translation, making robust malware detection feasible for low-resource ISAs.

\section{Model Design}


\subsection{Instruction Normalization} \label{subsec:normalization}

A binary, after being disassembled, is represented as a sequence of instructions. An instruction includes an opcode 
and zero or more operands.
We regard \emph{opcodes/operands as words} and \emph{basic blocks as sentences}. A basic block is a straight-line sequence of instructions with no branches inside it.

In Natural Language Processing (NLP), the out-of-vocabulary (OOV) issue is a well-known problem. To mitigate the OOV problem, we employ the normalization strategy.

\begin{itemize}
\aaf
  \item (R1): We use IDA Pro~\cite{IDA} to disassemble binaries, which generates dummy names~\cite{ida-dummy}. 
  We replace dummy names with their respective prefixes. 
  For example, \verb|off_| and \verb|seg_| represent offset pointer value and segment address value. They are replaced with \verb|<OFF>| and \verb|<SEG>|. 
    \aaf
\item (R2): Function names are replaced by \verb|<FUNC>|.

    \aaf
\item (R3): Number and hexadecimal constants are replaced by \verb|<VALUE>| and \verb|<HEX>|, respectively. 

\aaf
\end{itemize}

We provide examples in  Appendix~\ref{appendix-instruction-normalizing} that illustrate how these normalization rules are applied to assembly code across different ISAs.

\begin{figure}[t]
  \centering
  \includegraphics[width=1\linewidth]{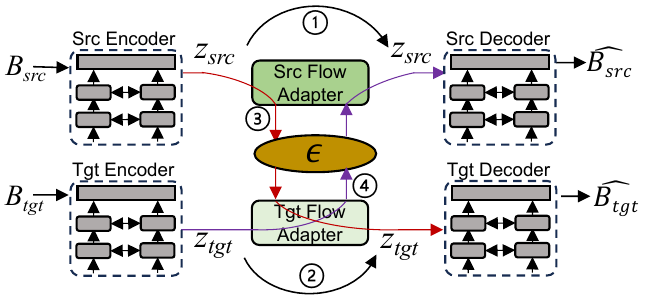}
  \aaf\af\aaf
  \caption{Model architecture of \toolmal, which contains a pair of flow adapters for modeling the distributions of basic block representations in the source ISA and target ISA, respectively.}
  \aaf
  \label{fig:nf-work_flow}
\end{figure}

\subsection{Normalizing Flows} \label{sec:flow}

We consider basic blocks as sentences, and instructions as words. The prior state-of-the-art, \tooluns, assumes that the latent representations of source and target basic blocks share a common semantic space, and learns ISA-agnostic latent representations, which abstract away ISA-specific details. However, this can be overly restrictive, as different ISAs exhibit distinct syntactic and structural characteristics that are not always easily aligned in a shared latent space. By enforcing ISA-agnostic representations, \tooluns\ may lose critical ISA-specific features necessary for accurate translation, leading to suboptimal performance. 

Our model \toolmal\ assumes the representations are different due to ISA-specific characteristics. Thus, they are modeled separately for each ISA, which makes it possible to better capture basic-block semantics in an ISA-specific manner. 
Specifically, \toolmal\ uses a pair of NFs (i.e., source flow and target flow) to model the distributions of basic-block latent representations in the source and target ISA. During translation, a latent code transformation is performed to transform the source representation $z_{src}$ into the target representation $z_{tgt}$, which is fed to the decoder to generate a better basic block in the target ISA. 
Figure~\ref{fig:nf-work_flow} shows how NFs are employed in our model.


 \noindent \textbf{Modeling Latent Representations by NFs.}  We use a pair of NFs to model the distribution of the basic block latent representations in different ISAs, i.e., $p_{z_{src}}(z_{src})$ and $p_{z_{tgt}}(z_{tgt})$, where $z_{src}$ and $z_{tgt}$ are the latent representation of the source basic block and target basic block, respectively.  Through NFs, we transform the distributions of the source and target representations to the base distribution $\epsilon$ (e.g., standard normal distribution), which can be viewed as the ``true'' underlying semantic space, abstracting away from ISA specifics.
We denote such mappings as $\textbf{G}_{(z_{src} \rightarrow \epsilon)}$ and $\textbf{G}_{(z_{tgt} \rightarrow \epsilon)}$.

\noindent \textbf{Designing NFs.} 
We use Real-valued Non-volume Preserving (RealNVP), a type of NFs designed for efficient density estimation and sampling~\cite{dinh2022density},  to build the source and target flow. 
%

Two key components of RealNVP are: \emph{Multi-scale Architecture} and \emph{Affine Coupling Layer}. 
With them, RealNVP can: (1) capture ISA-specific representations for each ISA at both local and global scales; (2) transform these encoded latent representations from one ISA to another invertibly; and  
(3) train using maximum likelihood estimation (MLE) with a tractable Jacobian determinant.

We construct \toolmal\ with two RealNVP models for the source flow and target flow. Each RealNVP model consists of three sequential flows to perform the latent code transformation. We explicitly model the source and target blocks with $K$ sequential flows (where $K=3$ in our case): 
\aaf
\begin{equation}
\footnotesize
    \label{equ:nf-src}
    p_{z_{src}}(z_{src})=p_{\varepsilon}(\varepsilon)\prod_{i=1}^K\left|\det \frac{\partial f_{src}^{(i)}(z^{(i)})}{\partial z^{(i)}}\right|^{-1}
\end{equation}
\aaf\aaf\aaf\aaf

\begin{equation}
\footnotesize
    \label{equ:nf-tgt}
    p_{z_{tgt}}(z_{tgt})=p_{\varepsilon}(\varepsilon)\prod_{i=1}^K\left|\det \frac{\partial f_{tgt}^{(i)}(z^{(i)})}{\partial z^{(i)}}\right|^{-1}
\end{equation}
where $p_{\varepsilon}(\varepsilon)$ is a base distribution. 
We select standard Gaussian distribution $\mathcal{N}(0,1)$ as our base distribution here. $f_{src}^{(i)}$ is the $i$-th transformations for the source blocks. $z^{(i)}$ is the intermediate variable where  $z^{(1)}=\varepsilon$ and $z^{(K)}=z_{src}$.

We denote the mapping process in Equation~(\ref{equ:nf-src}) as $\textbf{G}_{(\varepsilon \rightarrow z_{src})}$ and Equation~(\ref{equ:nf-tgt}) as $\textbf{G}_{(\varepsilon \rightarrow z_{tgt})}$. Given the invertibility feature of NFs, the mappings are also invertible. Thus, we have: $\textbf{G}_{(\varepsilon \rightarrow z_{src})}=\textbf{G}^{-1}_{(z_{src}\rightarrow \varepsilon )}$ and $\textbf{G}_{(\varepsilon \rightarrow z_{tgt})}=\textbf{G}^{-1}_{(z_{tgt} \rightarrow \varepsilon )}$. To achieve latent code transformation from source to target, we can formalize the transformation process as the composition of  $\textbf{G}_{(z_{src} \rightarrow \varepsilon)}$ and $\textbf{G}_{(\varepsilon \rightarrow z_{tgt})}$:

\begin{equation}
\footnotesize
    \textbf{G}_{(z_{src} \rightarrow z_{tgt})}=\textbf{G}_{(z_{src} \rightarrow \varepsilon)} \circ \textbf{G}_{(\varepsilon \rightarrow z_{tgt})}
\end{equation}

Therefore, we can learn a transformation $\textbf{G}_{(z_{src} \rightarrow z_{tgt})}$ by learning the transformation $\textbf{G}_{(z_{src} \rightarrow \varepsilon)}$ and $\textbf{G}_{(\varepsilon \rightarrow z_{tgt})}$. We also notice that $\textbf{G}_{(z_{src} \rightarrow z_{tgt})}$ and $\textbf{G}_{(z_{tgt} \rightarrow z_{src})}$ are invertible because they are compositions of two invertible mappings. Thus, 
We can follow a similar procedure and learn $\textbf{G}_{(z_{tgt} \rightarrow z_{src})}$.

\subsection{Encoder \& Decoder for Code Translation}



For binary code translation, different ISAs exhibit significant syntactic and semantic differences. To capture and align these variations, we propose a separated-encoder design, where each ISA is assigned a dedicated encoder. 
The source encoder and decoder are responsible for encoding and generating basic blocks in the source ISA, while the target encoder and decoder handle basic blocks in the target ISA.  \toolmal\ employs a multi-layer bi-directional Transformer~\cite{vaswani2017attention} to construct the encoders and decoders for the source and target ISA, as shown in Figure~\ref{fig:nf-work_flow}.



\noindent \textbf{Encoding}. 
Here we describe how the source encoder encodes the source basic block and generates the source basic block representation $z_{src}$. A similar procedure is applied to the target basic block representation generation.  

The source encoder processes the source basic block $\textbf{B}_{src}=\{\boldsymbol{b}_0, \cdots,\boldsymbol{b}_S\}$, and generates the hidden representations $\{\boldsymbol{h}_0, \cdots \boldsymbol{h}_S\}$. 
The basic block-level representation $z_{src}$ is computed using Equation~(\ref{equ:flow-encode}). The target encoder follows a similar process to encode the target basic block.

\aaf\aaf\aaf
\begin{align}
z_{src} = \boldsymbol{W}(
    &\ \text{max-pool}([\boldsymbol{h}_0, \dots, \boldsymbol{h}_S]) \notag \\
    &+ \text{mean-pool}([\boldsymbol{h}_0, \dots, \boldsymbol{h}_S]) \notag \\
    &+ \boldsymbol{h}_0)  \label{equ:flow-encode}
\end{align}


\noindent \textbf{Cross-lingual Translation.} To enable the decoder to better leverage ISA-specific latent representations, we perform a latent code transformation. For example, as shown in Figure~\ref{fig:nf-work_flow}, after calculating $z_{src}$ using Equation~(\ref{equ:flow-encode}), we employ the source flow to transform $z_{src}$ to $\epsilon$, and then the target flow to transform $\epsilon$ to $z_{tgt}$, which is in the target latent representation space. Then, the target decoder can decode $z_{tgt}$ to generate the target basic block. 

In contrast, the prior work \tooluns\ indiscriminately uses the same representational space for both source and target, without performing latent code transformation. 

\noindent \textbf{Decoding.} To capture the semantics and mitigate improper alignments between the source and target ISA, we follow the procedure in~\cite{setiawan2020variational} to generate the decoded representation: 
\aaf
\begin{equation}
    \label{equ:flow-decode}
    \boldsymbol{o}_i = (1 - \boldsymbol{g}_i) \odot \boldsymbol{s}_i + \boldsymbol{g}_i \odot \boldsymbol{z}
    \af
\end{equation}
where $\boldsymbol{g}_i=\sigma([\boldsymbol{s}_i;\boldsymbol{z}])$, $\sigma(\cdot)$ is the sigmoid function. $\boldsymbol{o}_i$ denotes Hadamard product between two tensors. The values in $\boldsymbol{g}_i$ control the contribution of $\boldsymbol{z}$ to $\boldsymbol{o}_i$.

\subsection{Model Training}
We first employ the causal language modeling (CLM) and masked language modeling (MLM) objectives to pretrain the encoders and decoders. 
We then train \toolmal\ in an unsupervised manner using three objectives: \emph{denoising auto-encoding (DAE)}, \emph{back translation (BT)}, and \emph{maximum likelihood estimation (MLE)}. The DAE reconstructs a basic block from its noised version, as illustrated in steps \circled{1} and \circled{2} in Figure~\ref{fig:nf-work_flow}. The BT process, shown in steps \circled{3} and \circled{4}, involves performing the latent code transformation twice while jointly training the encoders and decoders. For example, in BT for the source ISA, the transformation first occurs in the source-to-target direction, followed by the target-to-source direction. Appendix~\ref{appendix-training-details} provides further details. 

\noindent \textbf{{Maximum Likelihood Estimation (MLE)}.} The NFs are directly trained by maximum likelihood estimation (MLE) of basic-block level latent representations. The objective can be formulated as: 
\aaf
\begin{equation}
\footnotesize
    \label{equ:flow-mle1}
    \mathcal{L}_{MLE}(\textbf{G}_{(z_{src}\rightarrow \varepsilon)})=E_{z \sim p_{z_{src}}}[\log p_{z_{src}}(z)]
\end{equation}
where $E_{z \sim p_{z_{src}}}$ is approximated by sampling mini-batches of latent representations generated by the encoder during training. By minimizing the loss, 
we can construct $\textbf{G}_{(z_{src}\rightarrow z_{tgt})}$ and $\textbf{G}_{(z_{tgt}\rightarrow z_{src})}$.

Trained with the DAE, BT, and MLE objectives, \toolmal\ effectively translates basic blocks across ISAs. During testing, it translates each basic block of a given binary from the source ISA and then concatenates the translated blocks to form a translated binary for the target ISA.
\section{Evaluation}

\subsection{Experimental Settings} \label{subsec:settings}

We implement \toolmal\ using Transformer with $64$ hidden units, $4$ heads, ReLU activations, a dropout rate of $0.1$, and learned position embeddings. 
Appendix~\ref{maltrans-details}
presents the details. All the experiments were conducted on a computer with a $64$-bit 3.6 GHz Intel Core i9-CPU, a Nvidia GeForce RTX 4090 and 64GB RAM.

\noindent \textbf{Model Comparison.} 
We consider four \emph{baseline models} and one \emph{Same-ISA model} for comparison. 


\begin{itemize}
\aaf
\item \emph{Baseline Model 1:} \tooluns~\cite{ahmad-luo-2023-unsupervised} is the first and only existing work that focuses on binary code translation leveraging deep learning (Section~\ref{subsec:limit-prior}).

\aaf
\item \emph{{Baseline Model 2}:} \tooluni~\cite{wang-security23} also aims to resolve the data scarcity issue in malware detection. \tooluni\ learns transfer knowledge to enable the reuse of a model across ISAs. In contrast, we focus on binary code translation to enable model reuse.
 

\aaf
\item \emph{Baseline Model 3:} \toolcross~\cite{wang2024learning}, similar to \tooluni, aims to learn transfer knowledge to enable model reuse. 

\aaf
\item \emph{Baseline Model 4:} \emph{IR-based malware detection model}. Intermediate representation (IR) can abstract away ISA differences and represent binaries in a uniform style~\cite{vex-ir}. 

\aaf
\item \emph{Same-ISA model.}
We consider a model trained and tested on the \emph{same} ISA, \emph{without employing any translation}, as the Same-SIA model. 
As expected, this model, \emph{if trained with sufficient data}, is likely to outperform a model trained on one ISA and tested on another, representing the best-case scenario. 
\end{itemize}
\aaf

\af \subsection{Training \toolmal} \label{subsec:training}

We consider seven ISAs: X86-64, i386, ARM64, ARM32, MIPS32, PPC32 and M68K. We translate binaries from other ISAs to X86-64.


\noindent \textbf{Training Datasets.}
We collect various opensource programs, including \emph{openssl}, \emph{binutils}, \emph{findutils}, and \emph{libgpg-error}, which are widely used in prior binary analysis works~\cite{asm2vec-Ding,li2021palmtree}. 
We compile them for each ISA using \texttt{GCC} with different optimization levels (O0-O3) and disassemble the binaries using IDA Pro~\cite{IDA} to extract basic blocks, which are normalized and deduplicated.
Finally, we create a training dataset for each ISA.
The adequacy of these datasets is evaluated in Appendix~\ref{data-adequacy}.

\noindent \textbf{Byte Pair Encoding.} We use byte pair encoding (BPE) to process the datasets.
We set the BPE merge times based on the principles derived from empirical experiments and theoretical insights:

\begin{itemize}
\aaf
  \item The vocabulary size discrepancy between the source and target ISA should $<15\%$. A large vocabulary discrepancy can lead to an imbalanced learning problem, 
  resulting in inefficiencies and overfitting~\cite{gowda2020finding}.
  
\aaf


\item The vocabulary size of each ISA should $<12k$, to balance capturing word semantics with computational resource constraints. A large vocabulary size can negatively impact model performance due to increased complexity and sparse token distributions~\cite{jean2014using}.
\aaf
\end{itemize}

As shown in Table \ref{tab:bpe-merge-time2}, we can see that the vocabulary discrepancy within each pair is small, making them well-suited for training. Note that these principles are tailored to our specific scenarios.

\begin{table}[t]
\centering
\small
\caption{Vocabulary size, BPE merge times, and joint vocabulary size for each pair of ISAs. }
\aaf
\label{tab:bpe-merge-time2}
\resizebox{0.49\textwidth}{!}{%
\begin{tabular}{ccccc}
\hline
ISA Pair  & Vocab. Size & Vocab. Size  & Merge& Joint Vocab.  \\
(src $\leftrightarrow$ tgt) & (src) & (tgt) & Time & Size \\
\hline 
\hline
i386 $\leftrightarrow$ X86-64 & $7,135$ &  $7,104$ & $10,000$ & $9,688$ \\
ARM32 $\leftrightarrow$ X86-64 & $9,416$ & $9,236$ & $22,000$ & $17,262$ \\
ARM64 $\leftrightarrow$ X86-64 & $5,142$ & $4,455$ & $9,000$ & $7,104$ \\
MIPS32 $\leftrightarrow$ X86-64 & $10,995$ & $11,620$ & $14,000$ & $12,685$ \\
PPC32 $\leftrightarrow$ X86-64 & $8,691$ & $9,504$ & $17,000$ & $13,032$ \\
M68K $\leftrightarrow$ X86-64 & $7,148$ & $6,484$ & $9,000$ & $8,386$ \\
\hline 
\end{tabular}%
}
\aaf\af 
\end{table}

\noindent \textbf{Model Training.}
We first pre-train \toolmal\ and then
train it on the DAE, BT and MLE tasks
until the loss drops $<0.3$. 
The training time takes around $23h$, $22h$, $25h$, $23h$, $22h$ and $22h$, for i386$\leftrightarrow$X86-64, ARM32$\leftrightarrow$X86-64, ARM64$\leftrightarrow$X86-64, MIPS32$\leftrightarrow$X86-64, PPC32$\leftrightarrow$X86-64, and M68K$\leftrightarrow$X86-64, respectively.

\af \subsection{Testing on Binary Code Translation} \label{subsec:testing}

We use the Bilingual Evaluation Understudy (BLEU)~\cite{2002bleu} to evaluate the translation performance, which is often used to evaluate the quality of machine-generated translations by measuring the $n$-gram overlap between the translation and reference.  
We set the tokenization of \texttt{SacreBLEU}~\cite{sacrebleu} to \texttt{None}, apply add-one smoothing, and use the default settings 

\noindent \textbf{Testing Datasets.} 
We use three packages, \emph{zlib-1.2.11}, \emph{coreutils-9.0}, and \emph{diffutils-3.7}, to test \toolmal. Note that \emph{these packages are not included in the training dataset of \toolmal}. We compile them on the six ISAs using \texttt{GCC-11.4.0} with different optimization levels (O0-O3) and use IDA Pro to disassemble them. 

\noindent \textbf{Translation Results.}
We consider six ISAs, i386, ARM32, ARM64, MIPS32, PPC32 and M68K, as the source ISAs, and X86-64 as the target ISA. 
For each binary $B_1$ in the source ISA, there exists a semantically equivalent binary $B_2$ in X86-64.
We use \toolmal\ to translate $B_1$ into X86-64, resulting in a translated binary $B_3$ in X86-64. The BLEU score is computed between $B_3$ and $B_2$.
We report the average BLEU score for all binaries. The results are shown in Table \ref{tab:bleu-evaluation}. Appendix~\ref{appendix-translation} presents some translation examples. 

\begin{table}[t]
\centering
\small
\caption{BLEU scores of \toolmal\ \& baseline.}
\label{tab:bleu-evaluation}
\aaf
\resizebox{0.99\columnwidth}{!}{%
\begin{tabular}{ccc}
\hline
ISA Pair  & \toolmal\  & \tooluns\  \\

(src $\rightarrow$ tgt) &(\emph{our work}) & (\emph{baseline}) \\

\hline 
\hline
i386 $\rightarrow$ X86-64 & $0.41$ & $0.35$ \\
ARM32 $\rightarrow$ X86-64 & $0.40$ & $0.28$ \\
ARM64 $\rightarrow$ X86-64 & $0.37$ & $0.32$ \\
MIPS32 $\rightarrow$ X86-64  & $0.36$ & $0.27$ \\
PPC32 $\rightarrow$ X86-64  & $0.31$ & $0.25$ \\
M68K $\rightarrow$ X86-64 & $0.39$ & $0.31$ \\
\hline 
\end{tabular}%
}
\aaf\aaf
\end{table}

We compare \toolmal\ to the baseline \tooluns. We use the open-source trained model of \tooluns\ for this comparison. \textbf{\em Note that \tooluns\  focuses solely on two ISAs (X86-64 and ARM32). To ensure a comprehensive comparison, we train \tooluns\ on the same mono-architecture training datasets for the additional ISA pairs.}
We can see that \toolmal\ outperforms \tooluns\ across all ISA pairs and demonstrates satisfactory performance. Thus, \toolmal\ has good translation quality and can effectively translate binaries across ISAs. 

The average time to translate a basic block from one ISA to x86-64 is $10^{-4}$s, and the average number of blocks in a binary is $12,320$. Thus, translating a binary takes around $1.2$s. This demonstrates the good scalability of our approach.

\begin{table*}
\footnotesize
\caption{Malware detection results using LSTM. In Table (a), we compare the detection performance by translating malware using \toolmal\, \tooluns. Then we evaluate it against \tooluni,~\toolcross and the IR-based model. In Table (b), we give the results of the \emph{Same-ISA Model}, which is trained and tested on the \emph{same} ISA.}

\aaf\af
\label{tab:malware-result}
\begin{subtable}{0.72\textwidth}
\centering
\caption{\toolmal\ vs. Four baselines.}
\af
\resizebox{1.07\textwidth}{!}{%
\begin{tabular}{c|c|c|c|c|c}
\hline
ISA Pair & {\toolmal} & {\tooluns} &{\tooluni}& {\toolcross} &{IR-based Model} \\
(src $\rightarrow$ tgt) &(\emph{Our Work}) & (\emph{Baseline 1})  &  (\emph{Baseline 2})  &(\emph{Baseline 3}) &(\emph{Baseline 4}) \\
\hline 
\hline
i386 $\rightarrow$ X86-64 & \textbf{0.996} & $0.723$ & $0.953$  & - &$0.819$ \\
ARM32 $\rightarrow$ X86-64 & \textbf{0.981}  & $0.818$  &$0.953$ & $0.976$ &$0.815$ \\
ARM64 $\rightarrow$ X86-64 & \textbf{0.965}   & $0.638$ &$0.910$  & - &$0.825$ \\
MIPS32 $\rightarrow$ X86-64 & \textbf{0.971} & $0.725$ & $0.933$ & $0.964$ &$0.791$ \\
PPC32 $\rightarrow$ X86-64 & \textbf{0.973} & $0.722$ & $0.919$ & $0.967$ &$0.639$ \\
M68K $\rightarrow$ X86-64 & \textbf{0.931}   & $0.689$  &$0.893$  &  - &$0.476$ \\
\hline 
\end{tabular}
}
\end{subtable}
\qquad\quad
\begin{subtable}{0.21\textwidth}
\centering
\caption{The Same-ISA model.}
\af
\resizebox{0.97\textwidth}{!}{%
\begin{tabular}{c|cc}
\hline
Train \& Test on & Same-ISA\\
the \emph{same} ISA & Model \\
\hline 
\hline
i386  & $0.998$ \\
ARM32 & $0.986$ \\
ARM64 & $0.705$ \\
MIPS32 & $0.974$  \\
PPC32 & $0.849$  \\
M68K & $0.790$\\
\hline 
\end{tabular}
}
\end{subtable}
\aaf\af
\end{table*}

\subsection{Malware Detection Task} \label{subsec:malware-task}

We train a malware detection model on X86-64 (where abundant malware samples are available), and reuse the trained model for other ISAs by translating binaries using \toolmal~\footnote{We are aware that ARM32 is also a high-resource ISA; due to its importance, our evaluation involves it. However, whether our approach can be applied to translate other ISAs to ARM32 requires further experiments.}.




\noindent  \textbf{Malware Detection Model.} 
We use the LSTM model~\cite{lstm-2018} to detect malware, designed as two layers. The model details are provided in Appendix~\ref{lstm-details}. 
We first extract the token embeddings from \toolmal\ and integrate them into the input layer of LSTM. As a result, when a binary is fed into LSTM, each input token is represented as its embedding. 
To further enrich the task, we also apply a CNN model, with the results presented in Appendix~\ref{cnn-result}. In the following, we focus on the results related to LSTM.


\noindent  \textbf{Task-Specific \emph{Training} Datasets.} 
We collect  $2140$, $1740$, $1581$, $1430$, $962$, $545$, and $128$ malware samples from \emph{VirusShare.com}~\cite{VirusShare} for X86-64, i386, ARM32, MIPS32, PPC32, M68K, and ARM64, respectively. We spend significant efforts in collecting malware, especially for PPC32, M68K, and ARM64. Notably, our approach only needs malware in a high-resource ISA to train a malware detection model, significantly reducing the data collection burden for low-resource ISAs.

As we compare the results to the \emph{Same-ISA model} trained and tested on the \emph{same} ISA (Section~\ref{subsec:settings}), we build task-specific training datasets for each ISA. 
We split the malware samples into two parts: $80\%$ are used for training and $20\%$ for testing. Each training dataset also includes an equal number of benign samples, randomly selected from \emph{openssl}, \emph{binutils}, \emph{findutils}, and \emph{libgpg-error}.

We also evaluate the performance when testing on \emph{all} malware samples (without comparison to the \emph{Same-ISA model} baseline) in Appendix~\ref{all-testing}.

\noindent  \textbf{Task-Specific \emph{Testing} Datasets.} 
The testing dataset for each ISA includes equal numbers of malware and benign samples. 
The benign samples are randomly selected from \emph{zlib}, \emph{coreutils}, and \emph{diffutils}. 
Note that neither the benign nor malware samples are seen during the training of \toolmal\ or the malware detection model. 

%

\noindent \textbf{Result Analysis.} We first train LSTM on X86-64, and reuse it to test binaries in the other ISAs. The results in Table~\ref{tab:malware-result}(a) 
show when the model trained on X86-64 is transferred to i386, ARM32, ARM64, MIPS32, PPC32 and M68K, it achieves AUC = $0.996$, $0.981$, $0.965$, $0.971$, $0.973$ and $0.931$, respectively.
The high accuracies demonstrate the superior translation quality of \toolmal.

Next, we analyze how \toolmal\ preserves code semantics through translation.
We visualize the token embeddings from different ISAs and find that 
tokens performing similar operations, \emph{regardless of their ISAs}, have close embeddings. This demonstrates that \toolmal\ effectively captures code semantics across ISAs. Further details are provided in Appendix~\ref{appendix-semantic-analysis}.

\subsection{Model Comparison}

We compare \toolmal\ to the baselines and Same-ISA model (as described in Section~\ref{subsec:settings}).

\noindent \textbf{Comparison with Baseline Methods.} The first baseline is \tooluns.
\emph{As \tooluns\  focuses solely on two ISAs (X86-64 and ARM32), we train it on the same training datasets for the additional ISA pairs.}
We use it to translate binaries from the other ISAs to X86-64 and use LSTM to test the translated binaries.  The results in Table~\ref{tab:malware-result}(a) show that \toolmal\ achieves better translation quality, leading to improved malware detection performance. 

The second baseline is \tooluni~\cite{wang-security23}, which also aims to resolve the data scarcity issue in malware detection across ISAs. \tooluni\ learns transferable knowledge for model reuse. It covers four ISAs, including X86-64, ARM32, MIPS32, and PPC32.
To ensure a comprehensive comparison, we \emph{train it on the same datasets to cover the three additional ISAs (i386, ARM64 and M68K)}. Table~\ref{tab:malware-result}(a) show that \toolmal\ outperforms \tooluni\ across all ISA pairs.

The third baseline is \toolcross~\cite{wang2024learning}, which supports four ISAs (X86-64, ARM32, MIPS32, and PPC32). As it is not open-sourced, the comparison is limited to these four ISAs. The results show that \toolmal\ achieves better performance than \toolcross. 

The fourth baseline analyzes IR code. We assess whether a model trained on X86-64 IR code can be reused to test IR code in other ISAs. We use \texttt{angr}~\cite{vex-ir} to lift binaries into IR. We train LSTM using the same task-specific training dataset in X86-64, and apply the model to test the same task-specific datasets in other ISAs. The AUC scores are $0.819$, $0.815$, $0.825$, $0.791$, $0.639$ and $0.476$ for i386, ARM32, ARM64, MIPS32, PPC32 and M68K, respectively. This indicates IR alone does not magically allow a model trained on one ISA to be reused for other ISAs. (more explanation is discussed in Appendix~\ref{appendix-ir}).




\noindent \textbf{Comparison with Same-ISA Model.} The results of the Same-ISA model are shown in Table~\ref{tab:malware-result}(b). When the LSTM model is trained and tested on the \emph{same} ISA, it achieves AUC of $0.998$, $0.986$, $0.705$, $0.952$, $0.949$ and $0.790$ for i386, ARM32, ARM64, MIPS32, PPC32 and M68K, respectively. Comparing the results with our model in Table~\ref{tab:malware-result}(a), we observe: (1) our model achieves performance close to Same-ISA model for i386, ARM32 and MIPS32; (2) our model significantly outperforms Same-ISA model for ARM64, PPC32 and M68K.

For i386, ARM32, and MIPS32, the results align with expectations: the Same-ISA model, trained and tested on the \emph{same} ISA, outperforms \toolmal, which is trained on X86-64 and tested on other ISAs (via translation). 
However, for ARM64, PPC32, and M68K, our model outperforms the Same-ISA model due to the insufficient malware samples used to train the Same-ISA model. For example, for M68K, only $80\%$ of $245$  malware samples are used for training. This highlights the importance of sufficient training data to achieve desirable performance.
While including more data could enhance the model's performance, collecting malware samples for low-resource ISAs can be challenging. Our approach---translating binaries to x86-64---addresses data scarcity effectively. In an extreme case, \emph{if only one binary in a given ISA is available, we can still detect whether it is malware using the X86-64-trained model}.

\subsection{Hyperparameter and Ablation Study}

We conduct hyperparameter and ablation studies.

\noindent \textbf{Normalizing Flows.} We 1) explore different types of NFs, including \emph{scf} (Scaling and Coupling Flow) and \emph{glow} (Generative Flow), 2) vary the number of sequential flows, and 3) remove the flow adapter. The results show that 1) using \emph{scf} with 3 sequential flows yields the best performance, and 2) when the flow adapter is removed, the performance degrades, highlighting the crucial role of normalizing flows.

\noindent \textbf{Normalization Rules.} 
We conduct ablation study to evaluate the impacts of normalization rules (Section~\ref{subsec:normalization}). The results show that: (1) When all rules are applied, we achieve the best performance. (2) When a subset of rules is applied, the AUC values are reduced, indicating  each rule mitigates the OOV issue and positively impacts translation quality.



More details are presented in Appendix~\ref{appendix-hyperparameter-ablation-study}.

\section{Conclusion and Future Work}


We proposed \toolmal, a flow-adapter-based unsupervised binary code translation model. 
\toolmal\ incorporates normalizing flows (NFs) to model the basic block-level representations, enhancing its code translation capability.
We train \toolmal\ to translate binaries from other ISAs to X86-64, and reuse a malware detection model trained on X86-64 to test the translated code. 
Our approach effectively reduces the burden of data collection and achieves better malware detection across ISAs than prior state of the arts. 

Large language Models (LLMs) show extraordinary capability in coding tasks and are popular utilized~\cite{xu2025obliviate, ma2024one}. We leave using LLM for code translation as future work. 
\section*{Limitations}


\noindent \textbf{Generalizability.} Our evaluation encompasses malware detection across seven distinct ISAs, demonstrating the broad applicability of our method. However, our experimental validation is constrained to Linux-based malware samples. The transferability of our binary code translation framework to alternative operating systems—including Windows, iOS, and Android environments—remains an open question that warrants dedicated empirical investigation. Additionally, while our experiments focus on translating from various ISAs to X86-64, exploring ARM32 as an alternative target architecture presents an interesting direction, given its substantial malware dataset availability.

Beyond malware detection, our translation framework shows promise for broader binary analysis applications. The underlying code transformation capabilities could be leveraged for tasks such as vulnerability discovery and code similarity analysis. The strong performance demonstrated in malware detection validates the core technical approach, establishing a foundation for future research directions. Expanding this methodology to additional security analysis tasks represents a natural progression that could unlock significant research opportunities in cross-ISA binary analysis.

\noindent \textbf{Malware Packing and Obfuscation.}
The practice of malware packing entails hiding malicious code inside files that appear benign. Our research excludes malware packing from consideration, as it lies beyond our analytical scope. Our investigation centers on examining how well malware detection models transfer across different instruction set architectures (ISAs) through binary translation. The malware specimens employed in our detection experiments are uncompressed, enabling their examination and reverse engineering using analytical tools such as \texttt{IDA Pro}~\cite{IDA}.

Additionally, when dealing with compressed malware, one approach involves utilizing sophisticated decompression utilities, including PEiD~\cite{peid} and OllyDbg~\cite{OllyDbg}, to initially decompress the malware before analyzing the resulting content.

Code obfuscation methods present significant obstacles for malware identification systems. The malware specimens utilized in our research originated from \emph{VirusShare.com}~\cite{VirusShare}, a collection platform that gathers malware encountered in real-world environments. It is commonly understood that such malware frequently implements obfuscation strategies to circumvent detection systems. Nevertheless, we do not possess definitive information about the particular obfuscation methods employed in each malware sample, which complicates evaluating resistance against specific obfuscation approaches.

Our research primarily tackles the problem of limited data availability in resource-constrained ISAs by converting binaries from these architectures to a well-resourced ISA through \toolmal.

Subsequent research could methodically investigate how obfuscation techniques affect detection capabilities. A significant obstacle is the lack of a reliable, high-quality dataset that correlates malware samples with their corresponding obfuscation methods. Addressing this deficiency will be a priority in our upcoming investigations.
\section*{Ethical Considerations} 

To train and test our translation model \toolmal, we first collect various open-source programs and compile them for different ISAs using cross compilers. Given the wide availability of open-source programs, this requires minimal effort. Moreover, we strictly adhere to the licensing agreements and intellectual property rights associated with each program when collecting them and building the training and testing datasets. 

For evaluation within the security domain, we utilized malicious software specimens obtained from the \emph{VirusShare.com} platform~\cite{VirusShare}, a well-established digital repository that serves the cybersecurity research community. We acknowledge the substantial ethical responsibilities accompanying the use of such security-sensitive materials. Our research protocols emphasized legitimate scientific inquiry, prevention of inadvertent threat proliferation, and careful attention to confidentiality and regulatory compliance.

In the interest of scientific reproducibility and community advancement, we intend to distribute our developed model architecture, implementation code, and experimental datasets through a public repository under GPL licensing terms. The compiled open-source program collections will be made fully accessible in their original form. Regarding security-critical materials, we will exclusively share cryptographic hash identifiers and file references from \emph{VirusShare.com}, enabling verification and retrieval by qualified researchers while avoiding direct transmission of potentially dangerous executable content.

\newpage
\bibliography{custom}

\newpage
\appendix

\section{Appendix} \label{appendix}


\subsection{Model Training Details}
\label{appendix-training-details}
\noindent \textbf{Model Pretraining.}
Pretraining is a key ingredient of unsupervised neural machine translation. Studies have shown that the pre-trained cross-lingual word embeddings which are used to initialize the lookup table, have a significant impact on the performance of an unsupervised machine translation model~\cite{conneau2020unsupervised,conneau2019cross}. We adopt this and pre-train both the encoders and decoders of \toolmal\ to bootstrap the iterative process of our binary translation model. 
We employ causal language modeling (CLM) and masked language modeling (MLM) objectives to train the encoder and decoder.

(1) The CLM objective trains the model to predict a token $e_t$, given the previous $(t-1)$ tokens in a basic block $P(e_t|e_1,...,e_{t-1},\theta)$. (2) 
For MLM, we randomly sample $15\%$ of the tokens within the input block and replace them with \texttt{[MASK]} $80\%$ of the time, with a random token $10\%$ of the time, or leave them unchanged $10\%$ of the time. 

The first and last token of an input basic block is a special token \texttt{[/s]}, which marks the start and end of a basic block. We add position embedding and architecture embedding to token embedding, and use this combined vector as the input to the bi-directional Transformer network. Position embeddings represent different positions in a basic block, while architecture embeddings specify the architecture of a basic block. Both position and architecture embeddings are trained along with the token embeddings and help dynamically adjust the token embeddings based on their locations. 

\noindent \textbf{Denoising Auto-Encoding (DAE).} The DAE reconstructs a basic block from its noised version, as depicted in the process \circled{1} and \circled{2} in Figure~\ref{fig:nf-work_flow}. 
Given the input source block, $B_{src}$, we introduce random noise into it (e.g., altering the token order by making random swaps of tokens), resulting in the noised version, $B^{'}_{src}$. Then, $B^{'}_{src}$ is fed into the target encoder, whose output is analyzed by the source decoder. The training aims to optimize both the target encoder and source decoder to effectively recover $B_{src}$. Through this, the model can better accommodate the inherent token order divergences. 
Similarly, the source encoder and source decoder are optimized when the input is a target basic block, $B_{tgt}$. The training procedure of DAE involves only a single ISA at each time, without considering the final goal of translating across ISAs nor the latent code transformation. 

\noindent \textbf{Back Translation (BT).} We adopt the back-translation approach~\cite{feldman2020neural} to train our model in a translation setting, as shown in Figure~\ref{fig:nf-work_flow}. As an example,  given an input basic block in one ISA $N$, we use the model in the inference mode to translate it to the other ISA $M$ (i.e. applying the encoder of $N$ and the decoder of $M$). This way, we obtain a pseudo-parallel basic block pair. We then train the model to predict the original basic block (i.e. applying the encoder of $M$ and the decoder of $N$) from this synthetic translation.
As training progresses and the model improves, it will produce better synthetic basic block pairs through backtranslation, which will serve to further improve the model in the subsequent iterations.
In BT, the latent code transformation is performed twice and trained along with the encoders/decoders; taking BT for the source ISA as an example: first in the source-to-target direction, then in the target-to-source direction as shown in the steps of \circled{3} and \circled{4} in Figure~\ref{fig:nf-work_flow}.

\begin{table*}[ht]
    \centering
    \caption{Comparison of original and normalized assembly code.}
    \label{tab:instruction-normalizing}
    \scalebox{0.9}{
    \begin{tabular}{c c}
        \begin{subtable}{0.53\textwidth}
            \centering
            {\small
            \begin{tabular}{l@{\quad}|@{\quad}l}
                \texttt{\textbf{call} \_gpgrt\_log\_info} & \texttt{\textbf{call} <FUNC>} \\
                \texttt{\textbf{sbb} al 0} & 
                \texttt{\textbf{sbb} al <VALUE>} \\
                \texttt{\textbf{mov} esi 0ACh+\_bss\_start} & 
                \texttt{\textbf{mov} esi <HEX>+<TAG>}\\
                \texttt{\textbf{lea} rdi str\_LogWithPid} & \texttt{\textbf{lea} rdi <STR>}\\
                \texttt{\textbf{sub} rdx buffer} & 
                \texttt{\textbf{sub} rdx <TAG>}\\
            \end{tabular}}
            \caption{X86-64}
        \end{subtable} &
        \begin{subtable}{0.53\textwidth}
            \centering
            {\small
            \begin{tabular}{l@{\quad}|@{\quad}l}
                \texttt{\textbf{add} esp 0Ch} & 
                \texttt{\textbf{add} esp <HEX>} \\
                \texttt{\textbf{call} \_dcgettext} & \texttt{\textbf{call} <FUNC>} \\
                \texttt{\textbf{lea} eax [ebx-5D40h]} & \texttt{\textbf{lea} eax [ebx-<HEX>]}\\
                \texttt{\textbf{jmp} short loc\_37C3} & \texttt{\textbf{jmp} short <LOC>}\\
                \texttt{\textbf{mov} eax [150h+domainname] } & \texttt{\textbf{mov}  eax [<HEX>+<TAG>]}\\
            \end{tabular}}
            \caption{i386}
        \end{subtable} \\
        \begin{subtable}{ 0.53\textwidth}
            \centering
            {\small
            \begin{tabular}{l@{\quad}|@{\quad}l}
                \texttt{\textbf{ADD} R12 R12 0x1B000} & \texttt{\textbf{ADD} R12 R12 <HEX>} \\
                \texttt{\textbf{LDR} PC memcpy-0x2B7C} & \texttt{\textbf{LDR} PC <FUNC>-<HEX>} \\
                \texttt{\textbf{BEQ.W} loc\_109B4} & 
                \texttt{\textbf{BEQ.W} <LOC>}\\
                \texttt{\textbf{BL} gz\_uncompress} & \texttt{\textbf{BL} <FUNC>}\\
                \texttt{\textbf{CMP} R2 0} &
                \texttt{\textbf{CMP} R2 <VALUE>}\\
            \end{tabular}}
            \caption{ARM32}
        \end{subtable} &
        \begin{subtable}{ 0.53\textwidth}
            \centering
            {\small
            \begin{tabular}{l@{\quad}|@{\quad}l}
                \texttt{\textbf{LDR} X0 [SP \#0xC0+stream\_68]} & \texttt{\textbf{LDR} X0 [SP <HEX>+<TAG>]} \\
                \texttt{\textbf{TBZ} W0 \#0 loc\_AF38} & \texttt{\textbf{TBZ} W0 <HEX> <LOC>} \\
                \texttt{\textbf{ADRL} X1 str\_ErrorInitia} & 
                \texttt{\textbf{ADRL} X1 <STR>}\\
                \texttt{\textbf{B} \_gmon\_start\_} & \texttt{\textbf{B} <FUNC>}\\
                \texttt{\textbf{MOV} X19 \#0} & 
                \texttt{\textbf{MOV} X19 <HEX>}\\
            \end{tabular}}
            \caption{ARM64}
        \end{subtable} \\
        \begin{subtable}{ 0.53\textwidth}
            \centering
            {\small
            \begin{tabular}{l@{\quad}|@{\quad}l}
                \texttt{\textbf{lw} \$fp 0x40+var\_20} & \texttt{\textbf{lw} \$fp <HEX>+<VAR>} \\
                \texttt{\textbf{bal}   usage} & 
                \texttt{\textbf{bal} <FUNC>} \\
                \texttt{\textbf{beqz} \$v0 loc\_1358} &
                \texttt{\textbf{beqz} \$v0 <LOC>}\\
                \texttt{\textbf{move} \$a2 \$s3+1} & \texttt{\textbf{move} \$a2 \$s3+<VALUE>}\\
                \texttt{\textbf{sw} \$s0 0x40+path(\$sp)} & \texttt{\textbf{sw} \$s0 <HEX>+<TAG>(\$sp)}\\
            \end{tabular}}
            \caption{MIPS32}
        \end{subtable} &
        \begin{subtable}{ 0.53\textwidth}
            \centering
            {\small
            \begin{tabular}{l@{\quad}|@{\quad}l}
                \texttt{\textbf{addi} r3 r4 8 }& \texttt{\textbf{addi} r3 r4 <VALUE>} \\
                \texttt{\textbf{cmpwi} r2 8+var\_12} & \texttt{\textbf{cmpwi} <VALUE>+<VAR>} \\
                \texttt{\textbf{beq} \_str\_branch\_} & 
                \texttt{\textbf{beq} <TAG>}\\
                \texttt{\textbf{stw} r6 r2 6+0x1C5 } & \texttt{\textbf{stw} r6 r2 <VALUE>+<HEX> }\\
                \texttt{\textbf{lwz} r3 0x3H+str\_56} & \texttt{\textbf{lwz} r3  <HEX>+<STR>}\\
            \end{tabular}}
            \caption{PPC32}
        \end{subtable} \\
        \multicolumn{2}{c}{
        \begin{subtable}{0.53\textwidth}
            \centering
            {\small
            \begin{tabular}{l@{\quad}|@{\quad}l}
                \texttt{\textbf{move.l} \#\$DB3EF d0} & \texttt{\textbf{move.l} <HEX> d0} \\
                \texttt{\textbf{sub.w}   a2 d4-stream\_23} & 
                \texttt{\textbf{sub.w} a2 d4-<TAG>} \\
                \texttt{\textbf{tst.b} loc\_5H} &
                \texttt{\textbf{tst.b} <LOC>}\\
                \texttt{\textbf{cmp.l} \#\$352A \#23+mem\_name} & \texttt{\textbf{cmp.l} <HEX> <VALUE>+<TAG>}\\
                \texttt{\textbf{divs.w} \#16 d2} & \texttt{\textbf{divs.w} <VALUE> d2}\\
            \end{tabular}}
            \caption{M68K}
        \end{subtable}} \\
    \end{tabular}}

    \label{tab:asm_process}
\end{table*}

\subsection{Examples of Normalized Assembly Code} \label{appendix-instruction-normalizing}

Table~\ref{tab:instruction-normalizing} shows some randomly selected examples for instruction normalization. For each sub-table, the code on the left side represents the original version, while the code on the right side shows the normalized version.

\subsection{Dataset Adequacy} \label{data-adequacy}

To train \toolmal, we create a training dataset for each ISA: $2,789,119$ blocks for X86-64, $2,803,557$ blocks for i386, $7,413,083$ blocks for ARM64, $5,812,795$ blocks for ARM32, $4,813,685$ blocks for MIPS32, $3,964,237$ blocks for PPC32, and $5,463,395$ blocks for M68K. It should be noted that the training of \toolmal is unsupervised.

In NLP, it is widely recognized that a comprehensive dataset, which ensures that the vocabulary covers a wide range
of words, is crucial for training effective code translation models. We assess the adequacy of our mono-architecture datasets. Specifically, we study the vocabulary growth as we incrementally include programs. Our findings indicate that while the vocabulary size initially increases with the inclusion of more programs, it eventually stabilizes. Take x86-64 as an example, including \emph{openssl-1.1.1p} results in a vocabulary size of $23,029$. The size increases to $36,770$ (a $60\%$ growth) when \emph{binutils-2.34} is added, and then increases to $39,499$ (a $7.4\%$ growth) and $39,892$ (a $0.9\%$ growth) when \emph{findutils-4.8.0}, \emph{libpgp-error-1.45} are included, respectively. 
The growth trend is similar for other ISAs. It shows that the vocabulary barely grows in the end when more programs are added. According to the vocabulary growth trend and the high performance achieved, our mono-arch datasets are adequate to cover
instructions and enable effective code translation.

Note that the datasets used for training \toolmal\ have \emph{no overlap} with (1) the dataset used for testing the translation capability of \toolmal\ and (2) the testing dataset used in the malware detection task. The details of these datasets are introduced in the following sections.

\subsection{Model Parameters of \toolmal} 
\label{maltrans-details}

The encoders and decoders of \toolmal\ are implemented using the Transformer model. Table~\ref{tab:trans-params} shows the details of the model parameters. 

\begin{table*}[t]
\centering
\footnotesize
\caption{Parameter Details of \toolmal.}
\aaf
\label{tab:trans-params}
\resizebox{0.75\textwidth}{!}{
\begin{tabular}{ccc}
\hline
\textbf{Parameter} & \textbf{Value}    & \textbf{Description}                     \\ \hline \hline
Emb. Dimension     & 32/64/128         & Embedding layer size for tokens          \\
Hidden Dimension   & 4* Emb. Dimension & Transformer FFN hidden dimension         \\
Num. Layers        & 4                 & Number of transformer layers             \\
Num. Heads         & 4                 & Number of attention heads per layer      \\
Regu. Dropout      & 0.1               & Dropout rate for regularization          \\
Attn. Dropout      & 0.1               & Dropout rate in attention layers         \\
Batch Size         & 256               & Number of sentences per batch            \\
Max. Length        & 512               & Maximum length of one sentence after BPE \\
Optimizer          & Adam              & Adam optimizer with sqrt decay           \\
Clip Grad. Norm    & 5                 & Maximum gradient norm for clipping       \\
Act. Function      & ReLU              & Use ReLU for activation                  \\
Pooling            & Mean              & Use mean pooling for sentence embeddings \\
Accumulate Grad.   & 8                 & Accumulate gradients over N iterations   \\ \hline \hline

\end{tabular}
}
\end{table*}

\subsection{Model Parameters of Malware Detection LSTM Model} \label{lstm-details}

We use the LSTM model to detect malware. Table~\ref{tab:lstm-params} shows the parameters details of the LSTM model. 

\begin{table*}[t]
\centering
\footnotesize
\caption{Parameter Details of the malware detection LSTM model.}
\aaf
\label{tab:lstm-params}
\resizebox{0.75\textwidth}{!}{
\begin{tabular}{ccc}
\hline
\textbf{Parameter} & \textbf{Value} & \textbf{Description}                                   \\ \hline \hline
Emb. Dimension     & 32/64/128      & Input feature dimension for sequence embedding         \\
Num. of  Layers    & 3              & Number of stacked LSTM layers in the network           \\
Hidden Units       & 16             & Number of hidden units in each LSTM layer              \\
Output Units       & 1              & Dimension of the output layer                          \\
Batch Size         & 36             & Number of samples processed in one batch \\
Optimizer          & Adam           & Adaptive optimization algorithm with momentum          \\
Loss Function & BCEWithLogitsLoss & Binary cross-entropy with logits \\
Pooling     & Max    & Maximum value across temporal dimension                \\ \hline
\end{tabular}
}
\end{table*}

\begin{table*}[t!]
\centering
\footnotesize
\caption{Parameter Details of the malware detection CNN model.}
\aaf
\label{tab:cnn-params}
\resizebox{0.75\textwidth}{!}{
\begin{tabular}{ccc}
\hline
\textbf{Parameter} & \textbf{Value} & \textbf{Description}                                   \\ \hline \hline
Emb. Dimension     & 64      & Input feature dimension for sequence embedding         \\
Conv. Layers    & 2              & Number of convolutional layers in the CNN network           \\
Conv. Kernel       & 3             & Kernel size of the convolutional layers              \\
Output Units       & 1              & Dimension of the output layer                          \\
Batch Size         &64             & Number of samples processed in one batch \\
Optimizer          & Adam           & Adam optimizer with a learning rate of 0.001          \\
Loss Function & BCEWithLogitsLoss & Binary cross-entropy with logits \\
Pooling     & Max    & Maximum pooling to reduce spatial dimensions                \\ \hline
\end{tabular}
}
\end{table*}

\subsection{Malware Detection Using a CNN model}\label{cnn-result}

We use a 1-dimensional Convolutional Neural Network (CNN) model to detect malware. The parameters of the CNN model are presented in Table~\ref{tab:cnn-params}.

We use the same task-specific training and testing datasets described in Section~\ref{subsec:malware-task}. Moreover, we compare our results against four baseline methods and the Same-ISA model. Table~\ref{tab:cnn-malware-result} presents the malware detection results using the CNN model.


\begin{table*}
\footnotesize
\caption{Malware detection results using CNN. In Table (a), we compare the detection performance by translating malware using \toolmal, \tooluns\. Then we evaluate it against \tooluni\ and the IR-based and Same-ISA model. In Table (b), we give the results of the \emph{Same-ISA model}, which is trained and tested on the \emph{same} ISA.}
\af
\label{tab:cnn-malware-result}
\begin{subtable}{0.72\textwidth}
\centering
\caption{\toolmal\ vs. Four baselines.}
\resizebox{1.05\textwidth}{!}{%
\begin{tabular}{c|c|c|c|c|c}
\hline
ISA Pair & {\toolmal} & {\tooluns} &{\tooluni}& {\toolcross} &{IR-based Model} \\
(src $\rightarrow$ tgt) &(\emph{Our Work}) & (\emph{Baseline 1})  &  (\emph{Baseline 2})  &(\emph{Baseline 3}) &(\emph{Baseline 4}) \\
\hline 
\hline
i386 $\rightarrow$ X86-64 & \textbf{0.993} & $0.745$ & $0.953$  & - &$0.823$ \\
ARM32 $\rightarrow$ X86-64 & \textbf{0.984}  & $0.826$  &$0.953$ & $0.976$ &$0.842$ \\
ARM64 $\rightarrow$ X86-64 & \textbf{0.973}   & $0.641$ &$0.910$  & - &$0.818$ \\
MIPS32 $\rightarrow$ X86-64 & \textbf{0.974} & $0.762$ & $0.933$ & $0.964$ &$0.774$ \\
PPC32 $\rightarrow$ X86-64 & \textbf{0.972} & $0.736$ & $0.919$ & $0.967$ &$0.642$ \\
M68K $\rightarrow$ X86-64 & \textbf{0.931}   & $0.696$  &$0.893$  &  - &$0.483$ \\
\hline 
\end{tabular}
}
\end{subtable}
\qquad
\begin{subtable}{0.21\textwidth}
\centering
\caption{The Same-ISA model.}
\resizebox{0.88\textwidth}{!}{%
\begin{tabular}{c|cc}
\hline
Train \& Test on & Optimal\\
the \emph{same} ISA & Model \\
\hline 
\hline
i386  & $0.997$ \\
ARM32 & $0.988$ \\
ARM64 & $0.714$ \\
MIPS32 & $0.956$  \\
PPC32 & $0.853$  \\
M68K & $0.780$\\
\hline 
\end{tabular}
}
\end{subtable}
\end{table*}

When the CNN model trained on X86-64 is transferred to i386, ARM32, ARM64, MIPS32, PPC32 and M68K, it achieves AUC values of $0.993$, $0.984$, $0.973$, $0.952$, $0.942$ and $0.937$, respectively. These high accuracies highlight the superior translation quality of \toolmal, outperforming both \tooluns\ and the IR-based model. Compared to the Same-ISA model, we have the following observation. (1) First, our model achieves performance close to the Same-ISA model when testing malware on i386, ARM32, and MIPS32. This outcome is expected, as the Same-ISA model is trained and tested on the same ISA, while our model is trained on X86-64 and tested on other ISAs through translation.
(2) Second, our model significantly outperforms the Same-ISA model on ARM64, PPC, and M68K. This is due to the limited malware samples available for training the Same-ISA model on the two ISAs, highlighting the value of our approach. By reusing a model trained on a high-resource ISA, we enable robust detection for low-resource ISAs that would otherwise face significant challenges.

\subsection{Malware Detection Using All Malware Samples for Testing} \label{all-testing}

We train the LSTM model exclusively on X86-64, and reuse the trained model to test binaries in other ISAs, including i386, ARM32, ARM64, MIPS32, PPC32 and M68K.  It is important to note the key difference between the experiment described in this Appendix and that in Section~\ref{subsec:malware-task}. Here, we use all malware samples from i386, ARM32, ARM64, MIPS32, PPC32 and M68K for testing. In contrast, in Section~\ref{subsec:malware-task}, only 20\% of the malware samples from these ISAs are used for testing, as the remaining 80\% are reserved for training. 

\begin{table*}[h]
\footnotesize
\caption{Malware detection results. We use all the malware samples in i386, ARM32, ARM64, MIPS32, PPC32 and M68K for testing. We compare the detection performance by translating malware using \toolmal\ and \tooluns, and evaluate it against the IR-based model.}
\label{tab:malware-result-all}
\centering
\resizebox{0.8\textwidth}{!}{%
\begin{tabular}{c|c|c|c|c|c}
\hline
ISA Pair & {\toolmal} & {\tooluns} &{\tooluni}& {\toolcross}&{IR-based Model} \\
(src $\rightarrow$ tgt) &(\emph{Our Work}) & (\emph{Baseline 1})  &  (\emph{Baseline 2})  &(\emph{Baseline 3})&(\emph{Baseline 4}) \\
\hline 
\hline
i386 $\rightarrow$ X86-64 & \textbf{0.997} & $0.728$ & $0.949$ & -& $0.821$ \\
ARM32 $\rightarrow$ X86-64 & \textbf{0.982}  & $0.821$  &$0.951$ &$0.975$ & $0.813$ \\
ARM64 $\rightarrow$ X86-64 & \textbf{0.963}   & $0.641$ &$0.913$ & - &$0.822$ \\
MIPS32 $\rightarrow$ X86-64 & \textbf{0.974} & $0.731$ & $0.931$ & $0.968$ &$0.796$ \\
PPC32 $\rightarrow$ X86-64 & \textbf{0.978} & $0.732$ & $0.916$ & $0.973$&$0.641$ \\
M68K $\rightarrow$ X86-64 & \textbf{0.940}   & $0.682$  &$0.894$& - & $0.480$ \\
\hline 
\end{tabular}
}
\end{table*}

\noindent \textbf{Result Analysis.} We first train LSTM on X86-64, and reuse the model to test binaries in the other ISAs. The results are shown in Table~\ref{tab:malware-result-all}.  
We can see that when the model trained on X86-64 is transferred to i386, ARM32, ARM64, MIPS32, PPC32 and M68K, it achieves AUC = $0.997$, $0.982$, $0.963$, $0.974$, $0.978$ and $0.940$ respectively.
The high accuracies demonstrate the superior translation quality of \toolmal.

\subsection{Semantic Transfer Analysis}
\label{appendix-semantic-analysis}

\begin{figure}[t]
    \centering
    \includegraphics[width=0.9\linewidth]{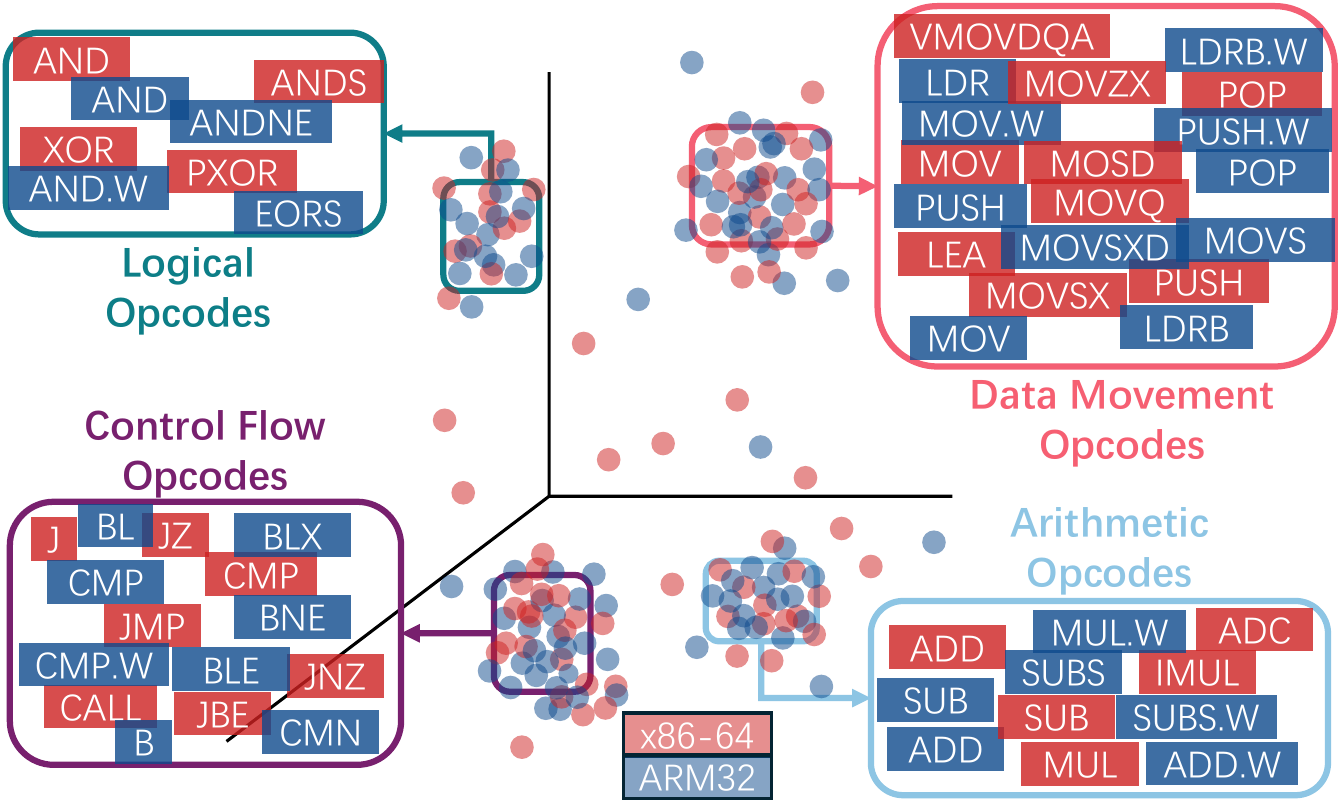}

    \caption{Visualization of opcode embeddings.}
    \label{fig:embedding}

\end{figure}
We analyze how \toolmal\ is able to preserve the code semantics through translation.
Specifically, we visualize the embeddings of opcode tokens from different ISAs. 
We take the X86-64 and ARM32 pair as an example. Opcodes, which determine the operation to be performed, capture more semantics compared to operands; we thus focus on opcodes for this demonstration. We extract the embeddings of $138$ X86-64 opcodes and $247$ ARM32 opcodes from \toolmal, and visualize them using t-SNE, as shown in Figure~\ref{fig:embedding}. Four categories of opcodes are selected for demonstration.
We can see that opcodes performing similar operations, \emph{regardless of their ISAs}, are close together. Thus, \toolmal\ can successfully capture semantic relationships of opcodes across ISAs and preserve code semantics.

\subsection{Hyperparameter and Ablation Study}
\label{appendix-hyperparameter-ablation-study}

\begin{table}[t]
\centering
\small
\caption{Impacts of different types of normalizing flows on the malware detection task.}
\label{tab:param-study-flow}
\resizebox{1\columnwidth}{!}{%
\begin{tabular}{c||c|c|c|c|c|c}
\hline
{Type}  & {i386} & {ARM32} & {ARM64} & {MIPS32} & {PPC32}&{M68K}  \\ 
  \hline \hline
3-scf  & \textbf{0.996}  & $0.981$ & \textbf{0.965} & \textbf{0.971} &\textbf{0.973}&\textbf{0.931}\\
3-glow & $0.994$  & \textbf{0.983} & $0.924$  & $0.823$ &  $0.873$& $0.873$\\
5-scf & $0.957$  & $0.875$ & $0.817$  & $0.910$ & $0.832$ &$0.824$\\
5-glow & $0.919$  & $0.891$ & $0.843$  & $0.804$ & $0.821$ &$0.760$\\
None & $0.854$  & $0.871$ & $0.784$  & $0.769$ & $0.811$ &$0.752$\\
\hline
\end{tabular}%
}
\end{table}

\noindent \textbf{Normalizing Flows.} 
We conduct an experiment to analyze the impact of normalizing flows on translation capability based on the following aspects: (1) exploring different flow-adapter architectures, such as Glow~\cite{diederik2014adam}; (2) varying the number of sequential flows to build the source/target flow; and (3) removing the flow-adapter architecture in \toolmal. 

The results are shown in Table~\ref{tab:param-study-flow}. We use two types of normalizing flows: \emph{scf} (Scaling and Coupling Flow) and \emph{glow} (Generative Flow), each configured with different numbers of sequential flows (e.g., 3 and 5). For instance, \emph{5-glow} represents the scenario where five consecutive \emph{glow} flows are utilized to construct the source/target flow. The last row, \emph{None}, represents the scenario in which no normalizing flows are incorporated into the model \toolmal.


We can observe that: (1) Using \emph{3-scf} to construct the normalizing flows yields the best performance across all ISAs except for ARM32. However, the result for ARM32 remains satisfactory, being close to the highest value. (2) When the flow-adapter architecture is removed from \toolmal, the performance degrades, highlighting the crucial role of normalizing flows in enhancing translation capability.

\noindent \textbf{Normalization Rules.}
Each instruction in the datasets is normalized by applying the three rules (R1, R2, and R3) discussed in Section~\ref{subsec:normalization}. Normalization is a vital step in our approach. In this experiment, we conduct ablation study by removing certain rules and evaluating their influence on malware detection. 
We consider these cases:
\begin{itemize}
\aaf\af
\item (\textbf{C1}): Applying all rules to the data.

\item (\textbf{C2}): Removing R1, applying R2 and R3.

\item (\textbf{C3}): Removing R2, applying R1 and R3.

\item (\textbf{C4}): Removing R3, applying R1 and R2.

\item (\textbf{C5}): Not applying any rules to the data.

\end{itemize}

\begin{table}[t]
\centering
\small
\af
\caption{AUC when varying normalizing strategies.}
\label{tab:param-study-preprocess}

\resizebox{1\columnwidth}{!}{%
\begin{tabular}{c||c|c|c|c|c|c}
\hline
{Case \#}  & {i386} & {ARM32} & {ARM64} & {MIPS32} & {PPC32}&{M68K}  \\ 
  \hline \hline
Case 1  & \textbf{0.996}  & \textbf{0.981} & \textbf{0.965} & \textbf{0.971} &\textbf{0.973}&\textbf{0.931}\\
Case 2 & $0.862$  & $0.521$ & $0.624$  & $0.623$ &  $0.617$&$0.638$\\
Case 3 & $0.883$  & $0.575$ & $0.617$  & $0.570$ & $0.503$ &$0.642$\\
Case 4 & $0.827$  & $0.533$ & $0.613$  & $0.422$ & $0.451$ &$0.546$\\
Case 5 & $0.793$  & $0.453$ & $0.523$  & $0.422$ &$0.473$ &$0.471$\\
\hline
\end{tabular}%
}
\end{table}
Table~\ref{tab:param-study-preprocess} shows the results.
We can observe that: (1) When all normalization rules are applied (\textbf{C1}), we achieve the best performance. (2) When a subset of normalization rules is applied (\textbf{C2-4}), the AUC values are lower than in \textbf{C1}, indicating that each normalization rule mitigates the OOV issue and has an impact on translation quality, thereby influencing malware detection performance. (3) When no normalization rule is applied (\textbf{C5}), the results are the lowest.

\noindent \textbf{Embedding Dimension.} We evaluate the impacts of the embedding dimension.
We test different dimension sizes, including $32$, $64$, and $128$, to train \toolmal. We then apply \toolmal\ to translate binaries from the source ISA to X86-64 for malware detection. 
The results are shown in Table~\ref{tab:parameter-emb-size}.
We observe that when the dimension is set to $64$, the AUC values are higher compared to other dimensions. Moreover, as the dimension increases, the training time also increases. We thus choose a dimension of $64$, considering both the translation quality and training efficiency. 

\begin{table}[t]
\centering
\small
\caption{AUC when varying dimension size.}
\label{tab:parameter-emb-size}
\resizebox{0.99\columnwidth}{!}{%
\begin{tabular}{cccc}
\hline
ISA Pair   & Dimension:  & Dimension:  & Dimension: \\  
(src $\rightarrow$ tgt)  & $32$& $64$ & $128$ \\
\hline
\hline
i386$\rightarrow$X86-64       & $0.995$ & $0.996$ & $0.994$ \\
ARM32$\rightarrow$X86-64      &  $0.978$ & $0.981$ & $0.972$ \\
ARM64$\rightarrow$X86-64      & $0.857$ & $0.965$ & $0.860$ \\
MIPS32$\rightarrow$X86-64      & $0.912$ & $0.971$ & $0.869$ \\
PPC32$\rightarrow$X86-64      & $0.917$ & $0.973$ & $0.892$ \\
M68K$\rightarrow$X86-64      & $0.832$ & $0.931$ & $0.917$ \\
 \hline
\end{tabular}%
}
\aaf\aaf\af
\end{table}

\subsection{Discussion on IR-based model}
\label{appendix-ir}

 Although IR can abstract away many architectural differences among ISAs, significant variations persist across IR code from different ISAs. Given two binary executables from different ISAs, compiled from the same source code, their corresponding IR representations can still appear quite different. This issue is discussed in Section 4.3.1 of the UniMap paper. Consequently, existing approaches that use IR for cross-ISA binary analysis often require additional advanced techniques, such as fuzzing or re-optimization, to bridge these differences. Similarly, in our evaluation, the IR-based baseline, which trains a model on x86 IR code and reuses it for testing on IR code from other ISAs, performs poorly due to these variations.  
 
Decompilation, which converts binary code into readable C-like code, is inherently a lossy process with limited accuracy (e.g., complex constructs such as optimized loops, switch statements, and inlined functions are often not recovered accurately). Moreover, while IDA Pro’s decompiler is among the best available, it supports only a limited set of ISAs, restricting its generalization to only those architectures. In contrast, our binary code translation approach does not suffer from these limitations. Thus, decompilation is an unsuitable solution for our needs.

\subsection{Translation Demonstration} \label{appendix-translation}

Table~\ref{tab:translation-demonstration} and Table~\ref{tab:translation-demonstration-2} shows some randomly selected examples. In each example, (1) the \emph{source ISA} is the original basic block in the source ISA, which could be i386, ARM32, ARM64, MIPS32, PPC32 or M68K; (2) the \emph{target ISA} is the basic block in the target ISA, X86-64, that is similar to the original basic block in the source ISA; and (3) the \emph{translated ISA} is the X86-64 basic block translated from the original basic block in the source ISA by our model \toolmal.

By comparing the translated X86-64 block with the target X86-64 block, we observe that \toolmal\ (1) accurately predicts almost all opcodes, and (2) while a few operands are predicted incorrectly, these errors are reasonable. Note that an instruction consists of an opcode (which specifies the operation) and zero or more operands (which specifies registers, memory locations, or literal data). Thus, opcodes, which determine the operation to be performed, capture more semantic information compared to operands. On the other hand, different registers or memory locations can store data while preserving code functionality, which reduces the significance of operands. 

Consider the first example of the ISA pair i386$\rightarrow$X86-64. In the target X86-64 basic block, the second instruction is: \texttt{\textbf{add} rbp [state+<HEX>]}, whereas in the translated X86-64 basic block, the predicted instruction is: \texttt{\textbf{add} rdx [s+<HEX>]}. Here, \toolmal\ successfully predicts the opcode \texttt{\textbf{add}}, and predicts a different register and memory cell for the operands, while preserving the functionality of the code.

\begin{table*}[h]
\centering
\caption{Examples for code translation.}
\label{tab:translation-demonstration}
\renewcommand{\arraystretch}{1.1}
\resizebox{1.98\columnwidth}{!}{%
\begin{tabular}{|c|c|c|c|}
\hline
      &      & Source i386    & \begin{tabular}[c]{@{}c@{}}\texttt{\textbf{sub} esp <VALUE> \textbf{add} edx [s+<HEX>] \textbf{mov} ebx esi \textbf{push} len \textbf{push} edx}\\ \texttt{\textbf{push} [esp+<HEX>+dictionary] \textbf{call} <FUNC> \textbf{add} esp <HEX> \textbf{jmp} <LOC>}\end{tabular} \\ \cline{3-4} 
      & 1 & Translated X86-64 & \begin{tabular}[c]{@{}c@{}}\texttt{\textbf{sub} rsp <VALUE> \textbf{add} rdx [s+<HEX>] \textbf{mov} rbx rbp \textbf{push} len}\\ \texttt{\textbf{mov} r8 [rsp+<HEX>+datalen] \textbf{mov} rcx len \textbf{call} <FUNC> \textbf{jmp} <ADDR>}\end{tabular}  \\ \cline{3-4} 
      &      & Target X86-64  & \begin{tabular}[c]{@{}c@{}}\texttt{\textbf{sub} rsp <VALUE> \textbf{add} rbp [state+<HEX>] \textbf{mov} eax <HEX> \textbf{push} <TAG>}\\\texttt{ \textbf{mov} [state+<HEX>] rdi \textbf{mov} rdi state \textbf{call} <FUNC> \textbf{jmp} short <TAG>}\end{tabular} \\ \cline{2-4} 
      &      & Source i386    & \begin{tabular}[c]{@{}c@{}}\texttt{\textbf{mov} edx [edi+<VALUE>] \textbf{mov} edp <VALUE> \textbf{cmp} edx <HEX> \textbf{lea} ecx [edx-<HEX>]}\\ \texttt{\textbf{setnz} al \textbf{and} ecx <HEX> \textbf{lea} ecx [edx-<HEX>]}\end{tabular} \\ \cline{3-4} 
i386  & 2 & Translated X86-64 & \begin{tabular}[c]{@{}c@{}}\texttt{\textbf{mov} rdx [abfd+<VALUE>] \textbf{mov} ebp <VALUE> \textbf{cmp} rdx <HEX> \textbf{lea} rcx [r1+rax+<VALUE>]}\\ \texttt{\textbf{setnz} al \textbf{and} ecx <HEX> \textbf{lea} rcx [r1+rax+<VALUE>]}\end{tabular} \\ \cline{3-4} 
      &      & Target X86-64  & \begin{tabular}[c]{@{}c@{}}\texttt{\textbf{mov} rdx rsp \textbf{mov} rsp <HEX> \textbf{cmp} rax rdx \textbf{lea} r8 <TAG>}\\\texttt{ \textbf{setnz} a1 \textbf{and} ecx <TAG> \textbf{lea} rcx [rax+<HEX>]}\end{tabular} \\ \cline{2-4} 
      &      & Source i386    & \begin{tabular}[c]{@{}c@{}}\texttt{\textbf{sub} esp <VALUE> \textbf{lea} eax [esi+<VALUE>] \textbf{push} eax \textbf{push} [esp+<HEX>+buf]} \\ \texttt{\textbf{call} <FUNC> \textbf{mov} edi eax \textbf{mov} c [esp+<HEX>+<TAG>]}\end{tabular} \\ \cline{3-4} 
      & 3 & Translated X86-64 & \begin{tabular}[c]{@{}c@{}}\texttt{\textbf{sub} rsp <VALUE> \textbf{lea} rax <TAG>+<HEX> \textbf{mov} cs:<TAG> rax \textbf{mov} rax [abfd+<VALUE>]}\\\texttt{ \textbf{call} <FUNC> \textbf{mov} rsi rsp \textbf{mov} rdi abfd \textbf{call} <FUNC> \textbf{add} rsp <VALUE>}\end{tabular} \\ \cline{3-4} 
      &      & Target X86-64 & \begin{tabular}[c]{@{}c@{}}\texttt{\textbf{sub} rsp <HEX> \textbf{lea} eax <HEX> \textbf{mov} rax rs:<HEX> \textbf{mov} [rsp+<HEX>+<VAR>]}\\ \texttt{\textbf{call} <FUNC> \textbf{mov} rcx rsp \textbf{mov} rsi <VALUE> \textbf{call} <FUNC> \textbf{add} eax <HEX>}\end{tabular}   \\ \hline
      &      & Source ARM32   & \begin{tabular}[c]{@{}c@{}}\texttt{\textbf{XOR} R0 R2 \textbf{MOV} R1 R1 <VALUE> \textbf{SUBS} R3 R0 R3 } \\\texttt{\textbf{LDP} R6 R6 have \textbf{BNE} <TAG> \textbf{CMP} copy R7 }\end{tabular} \\ \cline{3-4} 
      & 1 & Translated X86-64 & \begin{tabular}[c]{@{}c@{}}\texttt{\textbf{xor} edi edi \textbf{mov} rcx rdx \textbf{and} esi <VALUE> \textbf{sub} <TAG> <VALUE>}\\ \texttt{\textbf{mov} [rax+rdx-<VALUE>] cl \textbf{mov} ecx <VALUE>}\end{tabular} \\ \cline{3-4} 
      &      & Target X86-64  & \begin{tabular}[c]{@{}c@{}}\texttt{\textbf{xor} edi edi \textbf{mov} [rax+rcx-<VALUE>] di \textbf{and} edi <TAG> \textbf{sub} init curr}\\\texttt{ \textbf{mov} [s+<HEX>] rdx \textbf{mov} rsp <VALUE>}\end{tabular} \\ \cline{2-4} 
      &      & Source ARM32   & \begin{tabular}[c]{@{}c@{}}\texttt{\textbf{LDR} R2 [R3] \textbf{MOVS} R0 <VALUE> \textbf{BNE} <LOC>} \\\texttt{ \textbf{LDR} R0 [R2+<TAG>] \textbf{MOV} R1 <TAG> \textbf{BL} <FUNC>}\end{tabular} \\ \cline{3-4} 
ARM32 & 2 & Translated X86-64 & \begin{tabular}[c]{@{}c@{}}\texttt{\textbf{mov} rsi [rsp+<HEX>+p] \textbf{movsxd} rdi eax} \\\texttt{ \textbf{lea} rcx [rsp+<HEX>+id] \textbf{call} <FUNC>}\end{tabular} \\ \cline{3-4} 
      &      & Target X86-64  & \begin{tabular}[c]{@{}c@{}}\texttt{\textbf{mov} rdi in \textbf{movsxd} rdi file} \\ \texttt{\textbf{lea} rax [rsi+<VALUE>] \textbf{call} <FUNC>}\end{tabular} \\ \cline{2-4} 
      &      & Source ARM32   & \begin{tabular}[c]{@{}c@{}}\texttt{\textbf{MOV} R0 strm \textbf{MOVS} R1 <VALUE> \textbf{BL} <FUNC> \textbf{ADDS} err <VALUE>}\\ \texttt{\textbf{MOVS} R2 <TAG> \textbf{LDR} R1 <HEX> \textbf{BNE} <ADDR>}\end{tabular} \\ \cline{3-4} 
      & 3 & Translated X86-64 &  \begin{tabular}[c]{@{}c@{}}\texttt{\textbf{mov} edi ebx \textbf{lea} rcx <TAG> \textbf{movsxd} rax ds:<TAG> \textbf{call} <FUNC>}\\ \texttt{\textbf{add} rax rcx \textbf{lea} rdx <TAG> \textbf{jmp} <ADDR>}\end{tabular} \\ \cline{3-4} 
      &      & Target X86-64  & \begin{tabular}[c]{@{}c@{}}\texttt{\textbf{mov} edi ebx \textbf{lea} rsi <HEX> \textbf{movzx} rax <TAG> \textbf{call} <FUNC>}\\ \texttt{\textbf{add} eax ecx \textbf{lea} rbp <LOC> \textbf{jmp} <TAG>}\end{tabular} \\ \hline
      &      & Source ARM64   & \begin{tabular}[c]{@{}c@{}}\texttt{\textbf{MOV} W2 <VALUE> \textbf{MOV} W0 W2 \textbf{LDP} X2 <VALUE> [SP+<HEX>+<TAG>]}\\\texttt{ \textbf{LDR} X2 [SP+<HEX>+<TAG>] \textbf{LDP} X2 X3 [SP+<HEX>+<TAG>]}\end{tabular} \\ \cline{3-4} 
      & 1 & Translated X86-64 & \begin{tabular}[c]{@{}c@{}}\texttt{\textbf{mov} qword ptr [rax] <VALUE> \textbf{mov} rax [rbp+p]}\\\texttt{ \textbf{mov} rdi rax \textbf{call} <FUNC> \textbf{test} eax eax \textbf{jnz} <ADDR>}\end{tabular} \\ \cline{3-4} 
      &      & Target X86-64  & \begin{tabular}[c]{@{}c@{}}\texttt{\textbf{mov} qword ptr [rcx+<HEX>] <VALUE> \textbf{mov} [rdx+<HEX>] <VALUE>}\\ \texttt{\textbf{mov} [rdx+<HEX>] rcx \textbf{call} <FUNC> \textbf{test} rax rax \textbf{jnz} <TAG>}\end{tabular} \\ \cline{2-4} 
      &      & Source ARM64   & \begin{tabular}[c]{@{}c@{}}\texttt{\textbf{ADD} W2 W2 <VALUE> \textbf{ADD} X3 X0 X3 \textbf{SUB} W2 W2 <VALUE>}\\\texttt{ \textbf{MOV} W0 <VALUE> \textbf{STRB} W0 [X1+<HEX>] \textbf{MOV} X0 X2 <VALUE>}\end{tabular} \\ \cline{3-4} 
ARM64 & 2 & Translated X86-64 & \begin{tabular}[c]{@{}c@{}}\texttt{\textbf{mov} r1 q \textbf{mov} rsi p \textbf{movzx} ri byte ptr [p] }\\ \texttt{\textbf{sub} r1 <VALUE> \textbf{add} p <HEX> \textbf{mov} rax [p] \textbf{jmp} short <LOC>}\end{tabular} \\ \cline{3-4} 
      &      & Target X86-64  & \begin{tabular}[c]{@{}c@{}}\texttt{\textbf{mov} r14 rsp \textbf{mov} rsip \textbf{movzx} rax <HEX>} \\\texttt{ \textbf{sub} r10 <VALUE> \textbf{add} q <HEX> \textbf{mov} r11 <VALUE> \textbf{jmp} <TAG>}\end{tabular} \\ \cline{2-4} 
      &      & Source ARM64   &  \begin{tabular}[c]{@{}c@{}}\texttt{\textbf{LDR} W5 <VALUE> \textbf{LDR} X9 <HEX> \textbf{SUB} W0 W5 W0 \textbf{ADD} W1 W0 <VALUE> \textbf{LDP} W1 <VALUE>} \\ \texttt{\textbf{SUB} W0 W5 W0 \textbf{ADD} W1 W0 <VALUE> \textbf{LDP} W1 <VALUE> \textbf{LDRB} W2 <TAG>}\end{tabular} \\ \cline{3-4} 
      & 3 & Translated X86-64 & \begin{tabular}[c]{@{}c@{}}\texttt{\textbf{mov} rdx <HEX> \textbf{mov} [rsp+<HEX>+n] rdx \textbf{movzx} edx [r1+<HEX>] }\\ \texttt{\textbf{mov} esi edx \textbf{mov} r9 [r1+<HEX>] \textbf{lea} r1 [rdx+<TAG>]}\end{tabular} \\ \cline{3-4} 
      &      & Target X86-64  & \begin{tabular}[c]{@{}c@{}}\texttt{\textbf{mov} rdx [rbp+mode] \textbf{mov} ecx [rbp+fd] \textbf{movzx} rax [rbp+path] \textbf{mov} esi ecx}\\ \texttt{\textbf{mov} rdi rax \textbf{mov} [rbp+gz] rax \textbf{lea} r1 [rax+<VALUE>]}\end{tabular}  \\ \hline

\hline
  
\end{tabular}%
}
\end{table*}

\begin{table*}[h]
\centering
\caption{Examples for code translation.}
\label{tab:translation-demonstration-2}
\renewcommand{\arraystretch}{1.1}
\resizebox{1.98\columnwidth}{!}{%
\begin{tabular}{|c|c|c|c|}
\hline 

&      & Source MIPS32   & \begin{tabular}[c]{@{}c@{}}\texttt{\textbf{li} \$t9 <VALUE>   \textbf{lw} \$ra <HEX>+<VAR> (\$sp)    }\\\texttt{\textbf{addiu} \$t9 <FUNC>   \textbf{b} <FUNC>   \textbf{addiu} \$sp <HEX>     }\end{tabular} \\ \cline{3-4} 
      & 1 & Translated X86-64 & \begin{tabular}[c]{@{}c@{}}\texttt{\textbf{mov} r8 <VALUE>   \textbf{mov} rbx rsp+<HEX>+<TAG>  }\\\texttt{\textbf{lea} r10 r8+<ADDR>   \textbf{call} <FUNC>   add rbp <VALUE>   }\end{tabular} \\ \cline{3-4} 
      
      &      & Target X86-64  & \begin{tabular}[c]{@{}c@{}}\texttt{\textbf{mov} r10 <VALUE>  \textbf{mov} rax rsp+<HEX>+<VAR>  }\\ \texttt{\textbf{lea} r10 r10+<TAG>  \textbf{jmp} <FUNC>  \textbf{add} rsp <HEX>  }\end{tabular} \\ \cline{2-4} 

        &      & Source MIPS32   & \begin{tabular}[c]{@{}c@{}}\texttt{\textbf{li} \$a1 <VALUE>   \textbf{addiu} \$a1 <STR>   \textbf{move} \$a0 \$s0   }\\\texttt{\textbf{la} \$t9 <FUNC>   \textbf{jalr} \$t9   \textbf{lw} \$gp <HEX>+<VAR> (\$sp)  }\end{tabular} \\ \cline{3-4} 

MIPS32 & 2 & Translated X86-64 & \begin{tabular}[c]{@{}c@{}}\texttt{\textbf{mov} rsi <HEX>  \textbf{lea} rbp rbp+<ADDR>  \textbf{mov} rdi rax   }\\ \texttt{\textbf{mov} r11 <FUNC>   \textbf{call} <FUNC>   \textbf{mov} rap rsp   }\end{tabular} \\ \cline{3-4} 

      &      & Target X86-64  & \begin{tabular}[c]{@{}c@{}}\texttt{\textbf{mov} rsi <VALUE>  \textbf{lea} rsi rsi+<STR>  \textbf{mov} rdi, rbx} \\\texttt{\textbf{mov} r10 <ADDR>   \textbf{call} <FUNC>   \textbf{mov} rbp rsp}\end{tabular} \\ \cline{2-4} 
      
      &      & Source MIPS32   &  \begin{tabular}[c]{@{}c@{}}\texttt{\textbf{li} \$a2 <VALUE>   \textbf{li} \$a1 <VALUE>   \textbf{addiu} \$a1 <STR>   \textbf{move} \$a0 \$zero   \textbf{la} \$t9 <FUNC>   } \\ \texttt{\textbf{jalr} \$t9  lw \$gp <HEX+<VAR> (\$sp)   \textbf{move} \$a1 \$s1   \textbf{move} \$a0 \$v0 }\end{tabular} \\ \cline{3-4} 
      
      & 3 & Translated X86-64 & \begin{tabular}[c]{@{}c@{}}\texttt{\textbf{mov} rbx <HEX>  \textbf{mov} rsi <ADDR>  \textbf{ lea} rsi rsi+<TAG>   \textbf{xor} rsi rsi  }\\ \texttt{\textbf{mov} r10 r9 \textbf{call} <FUNC>  \textbf{mov} rbp rap \textbf{mov} rsi r14+<TAG>+<HEX> \textbf{mov} rdi rax  }\end{tabular} \\ \cline{3-4} 
      
      &      & Reference X86-64  & \begin{tabular}[c]{@{}c@{}}\texttt{\textbf{mov} rdx <VALUE>  \textbf{mov} rsi <VALUE>   \textbf{lea} rsi rsi+<STR>  \textbf{xor} rdi rdi }\\ \texttt{\textbf{mov} r10 
 r11 \textbf{call} <FUNC>  \textbf{mov} rbp rsp+<HEX>+<VAR> \textbf{mov} rsi r14 \textbf{mov} rdi rax}\end{tabular}  \\ \hline

    &      & Source PPC32   & \begin{tabular}[c]{@{}c@{}}\texttt{\textbf{lis} r13 r2@ha \textbf{addi} r3 r3 at1@l \textbf{lis} r9 <VAR>+<HEX>@ha  \textbf{lwz} r1 <TAG>+<LOC>@l(r9) }\\\texttt{ \textbf{add} r10 r7 r8 \textbf{add} r10 r6 r10 \textbf{lwz} r3 <HEX>(r1) \textbf{bl}  <FUNC>  }\end{tabular} \\ \cline{3-4} 
    
      & 1 & Translated X86-64 & \begin{tabular}[c]{@{}c@{}}\texttt{\textbf{mov} rbi rax \textbf{movsz} rbp [<HEX>+<VALUE>] \textbf{mov} rdx rax  \textbf{lea} [rax r1+<TAG>]}\\\texttt{\textbf{mov} rdx [rsi+<LOC>] \textbf{call} <FUNC> \textbf{test} rax rax \textbf{jz} <TAG> }\end{tabular} \\ \cline{3-4} 
      
      &      & Target X86-64  & \begin{tabular}[c]{@{}c@{}}\texttt{\textbf{mov} rbi <LOC> \textbf{mov} rbx <VALUE> \textbf{lea} [rdi+<LOC>]}\\ \texttt{\textbf{mov} rdx [rbx+<VALUE>] \textbf{call} <FUNC> \textbf{test} rdx rdx \textbf{jnz} <LOC>}\end{tabular} \\ \cline{2-4} 

        &      & Source PPC32   & \begin{tabular}[c]{@{}c@{}}\texttt{\textbf{mr} r3 r7 \textbf{li} r3 <VAR>  \textbf{lis} <FUNC>}\\\texttt{ \textbf{li} r6 <TAG> \textbf{mr} r5 r4 \textbf{stwu} r31 -4(r1) \textbf{b}  <LOC>}\end{tabular} \\ \cline{3-4} 

PPC32 & 2 & Translated X86-64 & \begin{tabular}[c]{@{}c@{}}\texttt{\textbf{movsx} r2 r13 \textbf{mov} rbi <VAR> \textbf{call} <FUNC>  }\\ \texttt{\textbf{mov} rbx <TAG> \textbf{mov} rax rdi \textbf{push} rbx  \textbf{jmp} <ADDR>}\end{tabular} \\ \cline{3-4} 

      &      & Target X86-64  & \begin{tabular}[c]{@{}c@{}}\texttt{\textbf{mov} r12 rbi \textbf{movsz} rsi \textbf{mov} rbx <LOC> \textbf{lea} <HEX>+<TAG>} \\\texttt{ \textbf{mov} r12 r7 \textbf{mov} rbx (<VALUE>+<HEX>) \textbf{mov} r8 <HEX> \textbf{jmp} <HEX>}\end{tabular} \\ \cline{2-4} 
      
      &      & Source PPC32   &  \begin{tabular}[c]{@{}c@{}}\texttt{\textbf{extsh} r2 r13 \textbf{li} r9 <VAR> \textbf{bl} <FUNC> \textbf{lis} r5 <TAG>@ha } \\ \texttt{ \textbf{addi} r5 r5 <TAG>@l \textbf{mr} r6 r3 \textbf{stwu} r5 -4(r1) \textbf{b} <ADDR>}\end{tabular} \\ \cline{3-4} 
      
      & 3 & Translated X86-64 & \begin{tabular}[c]{@{}c@{}}\texttt{\textbf{mov} rbx <LOC>+<HEX> \textbf{movsx} rbx rax \textbf{mov} eax [ecx+<HEX>] }\\ \texttt{\textbf{mov} ebi [ecx+<HEX>] \textbf{mov} r5 [r2+<VALUE>] \textbf{lea} r7 [rax+<HEX>]}\end{tabular} \\ \cline{3-4} 
      
      &      & Reference X86-64  & \begin{tabular}[c]{@{}c@{}}\texttt{\textbf{mov} rax [<VALUE>+<HEX>] \textbf{mov} eax <VAR> \textbf{mov} rbp [rsi+<VALUE>] \textbf{mov} eax ebx}\\ \texttt{\textbf{mov} rsi rdx \textbf{mov} rax (rax+<VAR>) \textbf{l} rbi [r3+<VAR>]}\end{tabular}  \\ \hline

    &      & Source M68K   & \begin{tabular}[c]{@{}c@{}}\texttt{\textbf{move.l} at2 a0  \textbf{move.l} <VALUE>+<TAG> a7 \textbf{move.l} a1 d2 \textbf{add.l} a2 d3 }\\\texttt{ \textbf{add.l} a3 d2 \textbf{move.l} <HEX>(a7) a0 \textbf{jsr} <FUNC> \textbf{tst.l} d0 \textbf{bne} <LOC> }\end{tabular} \\ \cline{3-4} 
      & 1 & Translated X86-64 & \begin{tabular}[c]{@{}c@{}}\texttt{\textbf{mov} rsi at3 \textbf{mov} rsp [<HEX>+<TAG>] \textbf{mov} rax rex  \textbf{lea} [r4 r10+r3]}\\\texttt{\textbf{mov} rbx [rbp+<VALUE>] \textbf{call} <FUNC> \textbf{test} rax rax \textbf{jz} <LOC> }\end{tabular} \\ \cline{3-4} 
      
      &      & Target X86-64  & \begin{tabular}[c]{@{}c@{}}\texttt{\textbf{mov} rsi <TAG> \textbf{mov} rbp <VALUE> \textbf{lea} [rdx+<HEX>]}\\ \texttt{\textbf{mov} rdi [rax+<VALUE>] \textbf{call} <FUNC> \textbf{test} rax rax \textbf{jnz} <TAG>}\end{tabular} \\ \cline{2-4} 

        &      & Source M68K   & \begin{tabular}[c]{@{}c@{}}\texttt{\textbf{move.l} d3 d2 \textbf{move.l} <HEX> a0 \textbf{lea} <FUNC> a1}\\\texttt{ \textbf{move.b} <TAG> d4 \textbf{ext.l} d4 \textbf{move.l} d1 d0 \textbf{move.l} a6 a7 \textbf{b} <LOC>}\end{tabular} \\ \cline{3-4} 

M68K & 2 & Translated X86-64 & \begin{tabular}[c]{@{}c@{}}\texttt{\textbf{mov} r12 rbx \textbf{mov} rdi <HEX> \textbf{lea} <FUNC> \textbf{movsx} rax <TAG> }\\ \texttt{ \textbf{mov} rbx rdx \textbf{push} rbp  \textbf{jmp} short <LOC>}\end{tabular} \\ \cline{3-4} 

      &      & Target X86-64  & \begin{tabular}[c]{@{}c@{}}\texttt{\textbf{mov} r12 rbp \textbf{mov} rsp \textbf{movsx} rcx <LOC> \textbf{lea} <TAG>} \\\texttt{ \textbf{mov} r2 r4 \textbf{mov} rbp (<TAG>+<HEX>) \textbf{mov} r11 <VALUE> \textbf{jmp} <ADDR>}\end{tabular} \\ \cline{2-4} 
      
      &      & Source M68K   &  \begin{tabular}[c]{@{}c@{}}\texttt{\textbf{move.l} <VALUE>+<VAR> d0 \textbf{move.l} d2 d1 \textbf{move.l} <VALUE>(d1) d2 \textbf{move.l} d1 a0  } \\ \texttt{ \textbf{} \textbf{move.s} d3 <VALUE> \textbf{move.l} <HEX>(a1) d3 \textbf{lea} <TAG>(a2) a1 }\end{tabular} \\ \cline{3-4} 
      
      & 3 & Translated X86-64 & \begin{tabular}[c]{@{}c@{}}\texttt{\textbf{mov} rax <VAR>+<HEX> \textbf{mov} rcx rbx \textbf{mov} ebx [ecx+<VALUE>] }\\ \texttt{\textbf{mov} ebi ecx \textbf{mov} r9 [r1+<HEX>] \textbf{lea} r1 [rdx+<TAG>]}\end{tabular} \\ \cline{3-4} 
      
      &      & Reference X86-64  & \begin{tabular}[c]{@{}c@{}}\texttt{\textbf{mov} rax [<VALUE>+<HEX>] \textbf{mov} eax <VAR> \textbf{mov} rbp [rsi+<VALUE>] \textbf{mov} eax ebx}\\ \texttt{\textbf{mov} rsi rbx \textbf{mov} rax (rax+<TAG>) \textbf{lea} rsi [r2+<VAR>]}\end{tabular}  \\ \hline

\end{tabular}%
}
\end{table*}

\subsection{Uniqueness of Our Work} \label{appendix-discussion}

We propose an entirely different approach compared to the prior works~\cite{wang2023can,wang2024learning}, such as \tooluni~\cite{wang2023can}, which also aims to reuse models across ISAs in order to resolve the data scarcity issue in low-resource ISAs.  

A key limitation of \tooluni~is its reliance on a linear transformation to map instruction embeddings between ISAs. This approach assumes that the embedding spaces of different ISAs exhibit structural similarity (i.e., are isomorphic). However, this assumption may not hold, particularly for ISAs with significantly different vocabulary sizes, making it infeasible to find a suitable linear transformation. In contrast, \toolmal~does not rely on such an assumption, making it more broadly applicable across diverse ISAs.

While \tooluni~learns cross-architecture instruction embeddings, our method focuses on translating binary code across ISAs. By translating code to a high-resource ISA, our approach offers several advantages. First, it allows the direct application of existing downstream models---trained on the high-resource ISA---to other ISAs through testing the translated code. In contrast, the works in~\cite{wang2023can, wang2024learning} require retraining the model using cross-architecture instruction embeddings. Furthermore, translating code from one ISA to another assists human analysts in understanding code from unfamiliar ISAs, supporting broader applications in code comprehension.

InnerEye~\cite{zuo2018neural} applies neural machine translation (NMT) techniques for binary code similarity comparison but \emph{does not perform binary code translation across ISAs}. It uses two encoders from NMT models, where each encoder generates an embedding for a piece of binary code of a given pairs, and measures similarity based on embedding distance. In contrast, our approach focuses on translating binary code across ISAs.

Intermediate Representation (IR) is an architecture-agnostic representation that abstracts away various architectural differences among ISAs. However, despite this abstraction, significant variations still exist across IR code from different ISAs. Consequently, even when binaries from different ISAs—compiled from the same source code—are converted into a common IR, the resulting IR code often differs significantly, as discussed in~\cite{ahmad-luo-2023-unsupervised}. We compare malware detection performance across ISAs by translating binaries using \toolmal\ against an approach that analyzes the IR code of binaries across ISAs. The results show our model enables superior cross-ISA malware detection (see the evaluation).
An interesting direction is exploring translation based on IR code, which we leave for future work.

\end{document}